\documentclass{vgtc}                          

\graphicspath{{figures/}{pictures/}{images/}{./}} 

\usepackage{times}                     

\usepackage{tabu}                      
\usepackage{booktabs}                  
\usepackage{lipsum}                    
\usepackage{mwe}                       

\usepackage{mathptmx}                  
\usepackage{graphicx}
\usepackage{subcaption}
\usepackage{flushend}
\onlineid{0}

\vgtccategory{Research}

\vgtcinsertpkg

\title{LLM-Based Authoring of Agent-Based Narratives through Scene Descriptions}

\author{Vinayak Regmi\thanks{e-mail: vregmi@purdue.edu} 
\and Christos Mousas\thanks{e-mail: cmousas@purdue.edu}} 
\affiliation{\scriptsize School of Applied and Creative Computing\\ Purdue University}

\abstract{
This paper presents a system for procedurally generating agent-based narratives using large language models (LLMs). Users could drag and drop multiple agents and objects into a scene, with each entity automatically assigned semantic metadata describing its identity, role, and potential interactions. The scene structure is then serialized into a natural language prompt and sent to an LLM, which returns a structured string describing a sequence of actions and interactions among agents and objects. The returned string encodes who performed which actions, when, and how. A custom parser interprets this string and triggers coordinated agent behaviors, animations, and interaction modules. The system supports agent-based scenes, dynamic object manipulation, and diverse interaction types. Designed for ease of use and rapid iteration, the system enables the generation of virtual agent activity suitable for prototyping agent narratives. The performance of the developed system was evaluated using four popular lightweight LLMs. Each model's process and response time were measured under multiple complexity scenarios. The collected data were analyzed to compare consistency across the examined scenarios and to highlight the relative efficiency and suitability of each model for procedural agent-based narratives generation. The results demonstrate that LLMs can reliably translate high-level scene descriptions into executable agent-based behaviors.
} 


\begin{document}


\firstsection{Introduction}

\maketitle

The rise of large language models (LLMs) has unlocked new creative opportunities in agent-based storytelling. LLMs, such as GPT-4, are trained on vast text corpora to learn statistical patterns in language and can generate coherent and contextually appropriate text in response to prompts. Their ability to perform in-context learning and plan complex behaviors has made them valuable tools for applications in creative writing, dialogue systems, and planning tasks~\cite{ref19}. Moreover, when paired with structured interfaces, LLMs can serve as tools for procedural content generation, capable of interpreting natural language input into structured outputs such as dialogue trees or command sequences~\cite{ref20}.

Building on these capabilities, procedural generation provides a complementary framework for translating high-level intent into executable content. In agent-based narrative contexts, procedural approaches model agents as goal-driven decision-makers whose interactions can generate plots, character arcs, and emergent group dynamics, supporting scalable simulations used in crowd modeling, training environments, and interactive media~\cite{ref22, ref27, ref28}. However, traditional agent-based systems rely on predefined timelines, behavior trees (BTs), or hard-coded logic, which are inflexible and time-consuming for authors and do not scale well to agent-based scenes or dynamic story generation~\cite{ref24, ref25, ref26}. These limitations motivate the need for systems that can translate high-level intent into executable behavior without manual scripting.

Thus, this paper presents a system for generating procedural agent-based narratives by combining LLM-based natural language interpretation with a modular agent behavior system. Users compose scenes by placing agents and objects in a 3D environment, after which the system dynamically generates a coherent narrative and visualizes it through animations and interaction tasks assigned to virtual agents by the LLM. Unlike traditional approaches, the proposed system leverages generative artificial intelligence (AI) to transform high-level prompts into low-level action sequences executed by virtual agents. The architecture supports modularity, extensibility, and ease of authoring, making it suitable for agent-based simulation and rapid agent-based narrative prototyping.

Specifically, the proposed system introduces a layered interaction model in which user-placed virtual agents and objects are parsed into a semantic scene description. A selected LLM transforms this description into behavior strings encoding movement, interaction type (e.g., grab, sit, touch), and timing. These behaviors are executed by our \texttt{SceneDirector} module, resulting in scenes that appear authored but are procedurally generated at runtime. This approach aligns with mixed-initiative authoring systems, where AI provides structure and coherence while users retain creative control~\cite{ref23}.

To evaluate the performance and practical suitability of large language models within the proposed system, four transformer-based models were integrated and tested: OpenAI's ChatGPT, Google's Gemini, Anthropic's Claude, and xAI's Grok. Each model was presented with semantically equivalent scene descriptions at controlled levels of complexity, ranging from simple single-agent interactions to multi-agent scenarios with compound object dependencies. Processing and response time were measured across repeated trials to assess statistical reliability, and average response values were computed to highlight performance differences, establishing a benchmark for lightweight LLMs in procedural generation contexts. Our system aims to support user-friendly experimentation in agent-based narratives. Additionally, the comparative analysis of transformer-based language models establishes a technical foundation for future research in AI-assisted narrative authoring.

\section{Related Work}
\label{sec2}

\subsection{Generative Agents and Narrative Behavior Systems}
\label{sec21}
Prior research has investigated how virtual agents can generate, represent, or enact narrative structures. Park et al.~\cite{ref6} simulated human-like behaviors and social dynamics in sandbox environments using layered memory, reflection, and planning mechanisms to produce emergent long-term behavior. Their work emphasized autonomous reasoning and internal state evolution rather than direct control over scene execution or animation. Ammanabrolu et al.~\cite{ref5} proposed a pipeline that expanded symbolic plot graphs into fluent narrative text, demonstrating how structured representations could be transformed into readable stories. Similarly, Qian et al.~\cite{ref7} explored collaborative agent-based planning in software engineering workflows, illustrating how modular agent architectures could coordinate complex task execution beyond narrative domains. Kybartas and Bidarra~\cite{ref2} categorized narrative generation systems into plot-based, character-based, and emergent approaches, highlighting challenges related to narrative coherence and the increasing role of learning-based techniques.

Behavior Trees (BTs) have also been widely used for authoring agent behavior in interactive narratives and games. Hu et al.~\cite{ref16} examined the scalability challenges of BTs in multi-agent narrative settings, noting that increasing scene complexity led to combinatorial growth in transitions and conflict-resolution logic. Kapadia et al.~\cite{ref17} introduced a flexible planning framework that allowed domain experts to define high-level behaviors while automatically generating actor trajectories within simulated environments. Their work emphasized the balance between authorial control and automated behavior synthesis in multi-actor systems.

\subsection{Multimodal and Visual Storytelling Frameworks}
\label{sec22}
A parallel body of work has explored multimodal storytelling systems that translate natural language prompts into visual or animated outputs~\cite{ref32}. Li et al.~\cite{ref10} presented a hierarchical LLM-based framework that converted short prompts into multimodal digital stories, including text, images, and animation assets. However, their system focused on offline content generation rather than interactive or real-time execution. Text-to-animation approaches have also been proposed to bridge language and motion. He et al.~\cite{ref14} demonstrated prompt-driven animation synthesis for virtual characters, while Liew et al.~\cite{ref4} performed temporally coherent video edits using latent diffusion techniques. Kapadia et al.~\cite{ref3} introduced a GUI-based authoring system that enabled users to construct narrative sequences using keyframes, timelines, and story arcs. While these frameworks produced visually compelling results, they generally relied on pre-authored animations, offline rendering pipelines, or fixed timelines. As such, they offered limited support for dynamic interaction, real-time character control, or environment-aware execution within a running simulation or game engine.

\subsection{Virtual Agent Interaction and Scene Affordance Models}
\label{sec23}
Research on virtual agent interaction has addressed how agents perceive, navigate, and manipulate objects within complex 3D environments. These approaches complement narrative and multimodal systems by focusing on physically grounded motion synthesis and interaction feasibility. Fraga et al.~\cite{ref29} modeled interaction dynamics using learned motion graphs combined with neural fields, enabling characters to perform context-appropriate motions that adapted to spatial constraints. Yu et al.~\cite{ref30} analyzed object interaction landscapes and proposed affordance-based representations that linked object geometry with semantically meaningful human poses. Their work improved an agent's ability to predict feasible interaction configurations relative to surrounding objects. Wu et al.~\cite{ref31} introduced a generative motion-planning framework for human-object interaction that combined learned representations with sampling-based planning. Their approach generated goal-directed interaction sequences that accounted for motion feasibility and environmental context. Although these systems primarily targeted motion quality and physical plausibility, they demonstrated techniques for translating high-level intent into structured, executable interaction steps.

\subsection{LLM-Driven Interaction Virtual Environments}
\label{sec24}
Large language models have increasingly been incorporated into virtual environments to enhance interactivity and responsiveness~\cite{ref34, ref35}. Normoyle et al.~\cite{ref9} employed GPT-3.5 to control expressive body language and emotional behaviors in non-player characters, extending beyond scripted dialogue systems. Li et al.~\cite{ref11} and Li et al.~\cite{ref13} explored adaptive storytelling mechanisms that used scene semantics or geographic context to influence narrative progression. These approaches tied story generation to environmental cues but emphasized context-aware narration rather than direct orchestration of character actions. Rychert et al.~\cite{ref15} integrated GPT into a VR escape-room experience, where the language model functioned primarily as a narrative guide. Ning and Pei~\cite{ref8} applied LLM-based reasoning to spatial rearrangement tasks focused on safety and accessibility rather than storytelling or agent behavior control. More recent work by Chang et al.~\cite{ref18} introduced semantic injection techniques to improve alignment between text descriptions and generated motion. While the aforementioned approaches targeted motion fidelity and token-level correspondence, they did not address higher-level interaction sequencing or multi-agent coordination within real-time environments.

\subsection{Contributions}
\label{sec25}
Prior research relevant to procedural storytelling and virtual agent behavior can be broadly grouped into four areas: (1) narrative logic and generative agent systems focused on symbolic planning, autonomy, and emergent behavior~\cite{ref2, ref5, ref6, ref7, ref16, ref17,ref33}; (2) multimodal and visual storytelling frameworks that translate prompts into visual sequences or animations using keyframes, diffusion models, or pre-authored motion synthesis~\cite{ref3, ref4, ref10, ref14, ref18}; (3) LLM-driven interaction and spatial storytelling systems in virtual environments~\cite{ref8, ref9, ref11, ref13, ref15}; and (4) motion- and interaction-centric methods that emphasize physical plausibility through learned interaction dynamics, affordance modeling, or generative motion planning~\cite{ref29, ref30, ref31}.

Building on these directions and unlike prior systems that rely on manually authored behavior trees, state machines, or animation timelines, this paper presents an integrated system for scene-aware agent-based action generation. The system allows users to arrange virtual agents and objects within a scene, automatically extract semantic metadata, and generate structured agent-based behavior plans using an LLM. These plans are encoded in a custom \texttt{SceneDirector} syntax and executed at runtime through coroutine-based logic that coordinates navigation, object interaction, and layered animation blending. By shifting authoring effort from low-level scripting to high-level scene composition, the approach supports rapid prototyping. It also enables non-technical users to design coherent, interactive agent-based narratives without extensive programming or animation expertise.

\section{Methodology}
\label{sec3}

\subsection{Methodology Research}
\label{sec31}
We introduce a system that combines LLM-driven natural language parsing with structured metadata assigned to virtual agents and objects in the scene. Our system produces agent-based sequences executed through animation and navigation systems. Our system leverages structured prompt design and scene-aware data serialization to generate interaction plans as logic strings. These are parsed into modular, coroutine-driven behaviors by our developed \texttt{SceneDirector} module.

\begin{figure*}[!htb]
 \centering 
 \includegraphics[width=\textwidth]{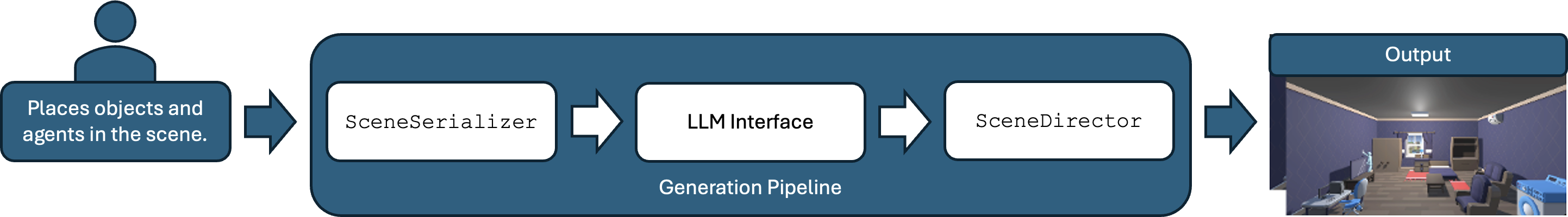}
 \caption{Overview of the LLM-driven interaction generation pipeline, from user-authored scene setup to the final execution.}
 \label{fig1}
\end{figure*}

\subsection{System Overview}
\label{sec32}
Our system leverages LLM-based logic generation and metadata-driven composition. It allows users to create complex, emergent story scenarios simply by placing virtual agents and objects into a scene and triggering the activity generation pipeline. The system comprises the following integrated components:
\begin{itemize}
\item \textbf{Scene Composition:} Users drag virtual agents and interactable objects (e.g., TVs, lamps, chairs) into the scene environment. Each object includes attached metadata via custom scriptable objects and \texttt{MonoBehaviours}.
\item \textbf{LLM Provider Selection:} Users select an LLM provider to generate interactions between agents and interactable objects. Our current implementation supports ChatGPT, Gemini, Claude, and Grok.
\item \textbf{Scene Metadata Serialization:} A \texttt{SceneSerializer} component extracts metadata from all agents and objects in the scene and formats it as a structured plain-language scene description. This includes agent IDs, object types, object interaction types, semantic tags, and the world-space positions of objects and agents.
\item \textbf{LLM Prompting:} The structured scene description is sent to the selected LLM via the API along with a predefined prompt. The prompt is designed to elicit structured responses in the \texttt{SceneDirector} format, a syntax that encodes the sequence of actions, interaction types, durations, and motion speeds for each agent.
\item \textbf{\texttt{SceneDirector} Parsing:} The \texttt{SceneDirector} module parses the LLM's output, maps it to in-scene object references, and assigns ordered \texttt{destination} queues to each agent. Each \texttt{destination} specifies the interaction type (i.e., \texttt{normal}, \texttt{grab}, \texttt{stationary}, or \texttt{basic}), the animation \texttt{duration}, and movement \texttt{speed}.
\item \textbf{Interaction Types:} The system supports four categories of agent-object interactions: \texttt{normal} interactions performed at a \texttt{destination}, \texttt{grab} interactions that involve object pickup and transport, \texttt{stationary} interactions that hold the agent in place for context-specific animations, and \texttt{basic} interactions that trigger simple object responses, such as toggling lights or activating devices.
\item \textbf{Agent Execution:} Each agent executes its interaction queue using pathfinding and animation playback. 
\item \textbf{Finalization:} After all destinations are executed, the agent returns to an idle animation state and stops pathfinding.
\end{itemize}

The system presents an end-to-end mechanism that converts scene composition into executable agent-based behaviors through structured LLM-generated instructions. Its modular design supports scalable interaction logic and flexible extension of narrative capabilities.

\subsection{Pipeline Overview}
\label{sec33}
The system generates procedural agent-based sequences by allowing users to assemble scenes from predefined agents and objects. The overall pipeline proceeds from scene composition to the execution of dynamic, LLM-driven animation behaviors. Figure~\ref{fig1} illustrates the end-to-end pipeline in which user-placed agents and objects are serialized into a plain-language scene description, sent through an LLM interface, and converted into a structured \texttt{SceneDirector} instruction string. This output is then parsed and executed in our scene environment, producing agent-based interactions based on the generated action plan.

\subsubsection{Scene Composition}
\label{sec331}
\begin{figure}[!htb]
    \centering
    \begin{subfigure}[b]{\columnwidth}
        \includegraphics[width=\textwidth]{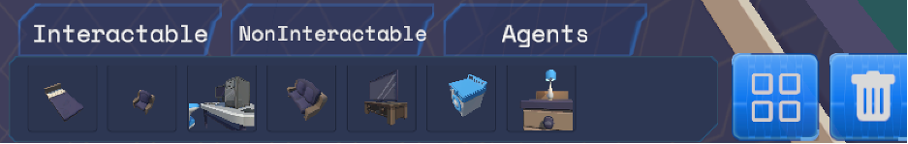}
        \caption{}
        \label{fig:gull}
    \end{subfigure}
    \begin{subfigure}[b]{\columnwidth}
        \includegraphics[width=\textwidth]{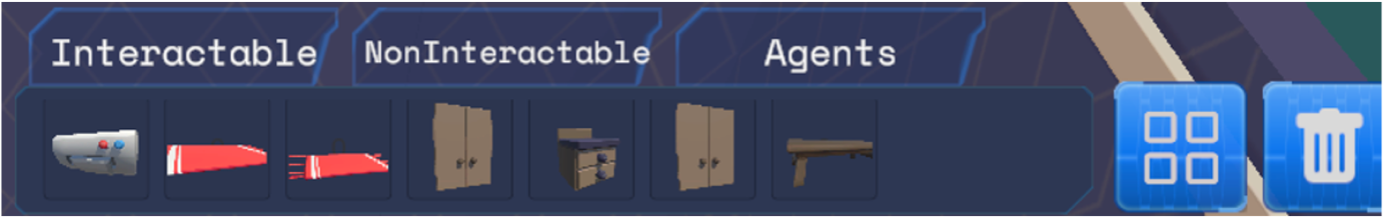}
        \caption{}
        \label{fig:tiger}
    \end{subfigure}
    \begin{subfigure}[b]{\columnwidth}
        \includegraphics[width=\textwidth]{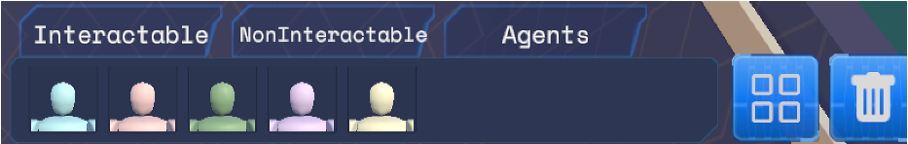}
        \caption{}
        \label{fig:mouse}
    \end{subfigure}
    \caption{Placable objects and agents selection menus: (a) interactable object menu, (b) non-interactable object menu, and (c) agent menu.}
    \label{fig2}
\end{figure}

Scene setup begins with the placement of prefabricated elements drawn from the object directory. These include virtual agents equipped with navigation, animation, and behavior components; interactable objects that respond to animations or context-specific actions, such as sitting, grabbing, or activating devices; and non-interactable objects that serve purely as environmental or decorative elements within the scene. These objects are placed into the grid system by selecting them from a menu, as shown in Figure~\ref{fig2}.

\begin{figure}[!htb]
 \centering 
 \includegraphics[width=\columnwidth]{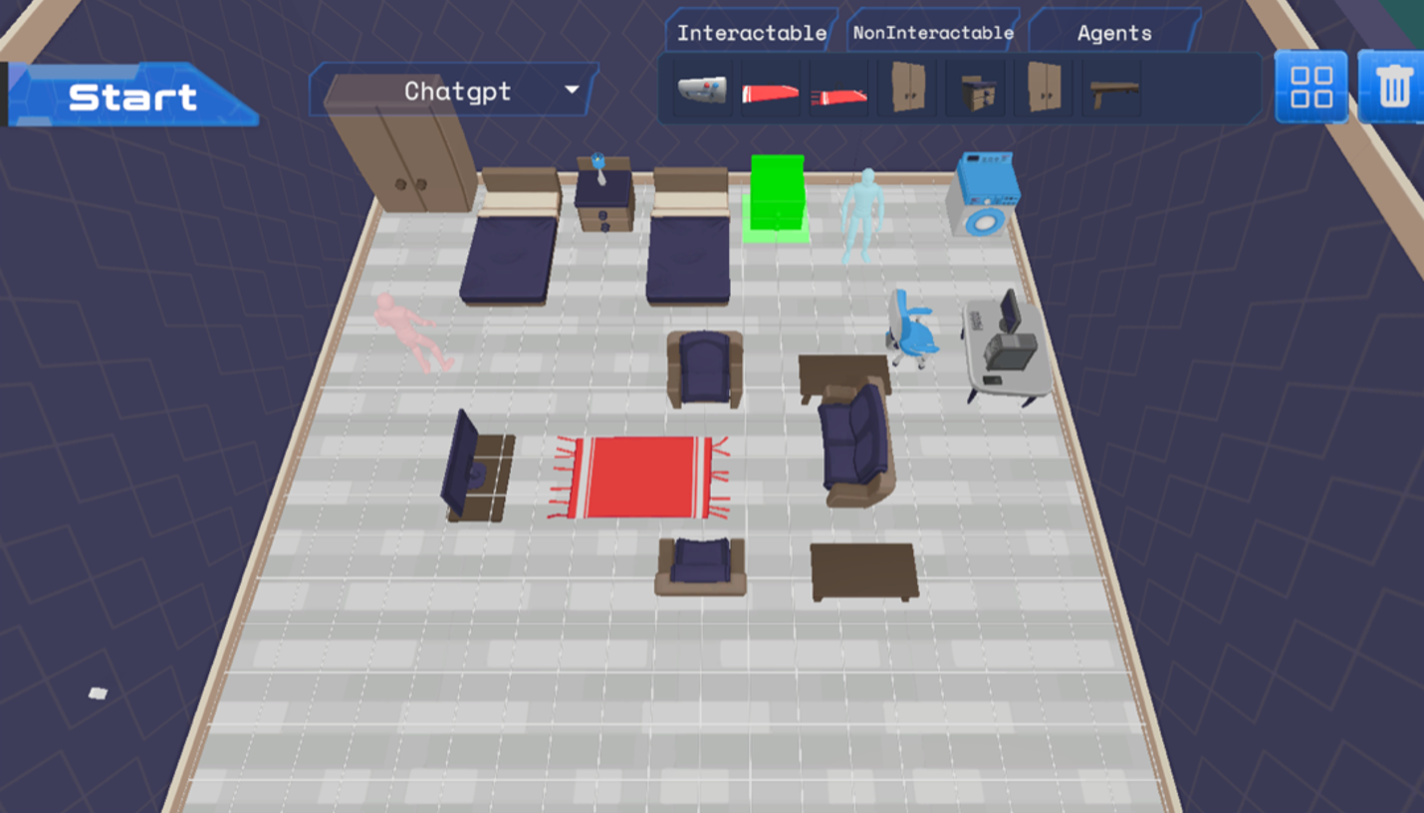}
 \caption{An example of a user-assembled scene including a virtual agent and several interactable and non-interactable (decorative) objects.}
 \label{fig3}
\end{figure}

Each placed virtual agent or object is associated with a corresponding metadata wrapper that encodes semantic and functional information. Virtual agents are assigned a \texttt{SelfExplainer} component that specifies attributes such as name, identifier, position, and semantic tags (e.g., \texttt{child}, \texttt{worker}). Interactable objects are linked to an \texttt{InteractableObjectExplainer} \texttt{ScriptableObject} that records the object's identifier, animation set, supported interaction types (i.e., \texttt{normal}, \texttt{grab}, \texttt{stationary}, and \texttt{basic}), semantic descriptors, and spatial position. This metadata forms the basis for later serialization and LLM-based behavior generation. An example of an assembled scene is shown in Figure~\ref{fig3}. Finally, once an object is placed in the scene, it can be repositioned, rotated, or removed through the object editing interface.

\subsubsection{LLM Provider Selection}
\label{sec332}
The user can select an LLM provider. In our current implementation, we included ChatGPT (gpt-4.1-mini), Claude (claude-sonnet-4-5), Gemini (gemini-2.5-flash), and Grok (grok-4-1-fast). The selected provider is used when creating the interactions. The scene metadata (see Section~\ref{sec333}) is sent to the selected LLM, which returns the interaction string (see Section~\ref{sec334}) that the \texttt{SceneDirector} can parse. All the LLMs are conditioned on the same system prompt, as shown in Table~\ref{tab1}, to tailor their responses to our application's needs.

\begin{table*}[!htb]
\caption{The system prompt that informs the chosen LLM on how to process the input information and generate the output scenario.}
\label{tab1}
\centering
\begin{tabular}{p{17cm}}
\toprule
You are a procedural story generation assistant. Your task is to convert a formatted scene description into a compatible \texttt{SceneDirector} instruction string used in our system. You should create a short, coherent story using the provided virtual agents and interactable objects.
\begin{itemize}
    \item Grabbed objects are destroyed after use and cannot be reused. 
    \item \texttt{grab} and \texttt{stationary} interactions are compatible and may be layered in sequence (e.g., carrying an object and sitting down).
    \item For all other interactions, objects must be freed before another agent may use them. 
    \item Overestimate occupancy: Account for time to walk to the object and interaction duration when tracking object availability.
    \item \texttt{speed} can range from 1.0 to 4.0. 
    \item \texttt{duration} can range from 2 to 16 seconds, so use the length that fits your storytelling.
    \item For \texttt{basic} interaction, keep the max limit at 5.00 and the min at 3.00.
\end{itemize}

Each virtual agent must be represented as: \texttt{AGENT\_ID \{ObjectID\_1 (T/F, DURATION, SPEED, GRAB\_TF, STATIONARY\_TF, BASIC\_TF), ObjectID\_2 (...)\}}, where:
\begin{itemize}
    \item \texttt{interact} (True/False): always True if any of grab, stationary.
    \item \texttt{duration}: interaction duration (e.g., 3.5 seconds).
    \item \texttt{speed}: movement speed (e.g., 1.0).
    \item \texttt{grab}: True if the agent picks up and carries the object.
    \item \texttt{stationary}: True if the agent performs a stationary action like typing, sitting, or sleeping.
    \item \texttt{basic}: True for short IK-based interactions (e.g., flipping a switch, turning on a light, or operating a machine).
\end{itemize}

Output **only** the \texttt{SceneDirector} string and nothing else. Example Output:
\begin{itemize}
    \item If \texttt{A\_1} turns on a light switch (\texttt{Obj\_1}), then sits at a desk (\texttt{Obj\_2}), your response should be: \texttt{A\_1 \{Obj\_1 (T, 2, 1, F, F, T), Obj\_2 (T, 5, 1, F, T, F)\}}.
\end{itemize}

You will be sent:
\begin{itemize}
    \item A plain-language scene description.
    \item A list of available agents and objects (with IDs and features).
    \item Your response must only contain the properly formatted \texttt{SceneDirector} string.
\end{itemize}

Note: There may be duplicate objects e.g. multiple beds, that is so that multiple agents can interact with different ones, make it make sense story wise, one agent may be able to interact with multiple light switches, but it does not make sense if an agent sleeps on one bed and gets up and sleeps on the other so be reasonable with the story and pay attention to these scenarios. Also, you can move to an object and not interact with it, as the output format suggests; do not always opt for this, but in a multi-agent scenario, if all objects are occupied, you may opt for this.\\
~\\
Remember to place a comma after each agent's complete entry before starting the next agent's entry, e.g., A\_1 \{...\}, A\_2 \{...\}.\\
~\\
Also, remember not to overlap actions between two agents, e.g., sitting on a couch while the other is still sitting, or interacting with a light switch while the other is doing the same.\\
~\\
Scene Note: All Lights are currently off; take that into consideration. You can turn them on. You can perform other activities and later turn them off.\\
~\\
Final Tip: Your main objective is to create a story and play it out based on what you have, so if some objects are not needed in your story, you do not always have to use them; just generate a relevant story.\\
\bottomrule
\end{tabular}
\end{table*}

\subsubsection{Scene Metadata Generation}
\label{sec333}
Once the scene is populated, a custom \texttt{SceneSerializer} script is triggered. This system scans the scene to collect metadata from virtual agents and interactable objects, then converts it into a structured string. This plain language ensures the LLM can interpret the scene semantically before generating structured behavior. Table~\ref{tab2} provided an example of a serialized scene description.

\begin{table*}[!htb]
\caption{Example of a serialized scene description.}
\label{tab2}
\centering
\begin{tabular}{ll}
\toprule
\texttt{1.}&  \texttt{Scene Description:}  \\
\texttt{2}.&  \texttt{----------} \\
\texttt{3}.&  \texttt{Actors:} \\
\texttt{4}.&  \texttt{----------} \\
\texttt{5}.&  \texttt{Name: Guy} \\
\texttt{6}.&  \texttt{ID: A\_1} \\
\texttt{7}.&  \texttt{Tags: male, college student, casual, claustrophobic} \\
\texttt{8}.&  \texttt{Position: (-.36, .11, -6.12)} \\
\texttt{9}.&  \texttt{----------} \\
\texttt{10}.& \texttt{----------} \\
\texttt{11}.& \texttt{Interactable Objects:} \\
\texttt{12}.& \texttt{----------} \\
\texttt{13}.& \texttt{Object ID: Obj\_5} \\
\texttt{14}.& \texttt{Name: Chair} \\
\texttt{15}.& \texttt{Is Grabbable: No} \\
\texttt{16}.& \texttt{Is Stationary: Yes} \\
\texttt{17}.& \texttt{Is Stationary Compatible: No} \\
\texttt{18}.& \texttt{Is Basic Interaction: No} \\
\texttt{19}.& \texttt{Tags: chair, sit, stay, relax} \\
\texttt{20}.& \texttt{Position: (-1.18, .23, -5.55)} \\
\texttt{21}.& \texttt{----------} \\
\texttt{22}.& \texttt{Object ID: Obj\_1} \\
\texttt{23}.& \texttt{Name: Computer} \\
\texttt{24}.& \texttt{Is Grabbable: No} \\
\texttt{25}.& \texttt{Is Stationary: Yes} \\
\texttt{26.}& \texttt{Is Stationary Compatible: No} \\
\texttt{27.}& \texttt{Is Basic Interaction: No} \\
\texttt{28.}& \texttt{Tags: work, play games, desktop, office work} \\
\texttt{29.}& \texttt{Position: (.70, .26, -5.46)} \\
\texttt{30.}& \texttt{----------} \\
\texttt{31.}& \texttt{----------} \\
\texttt{32.}& \texttt{END} \\
\texttt{33.}& \texttt{----------} \\
\bottomrule
\end{tabular}
\end{table*}

\subsubsection{LLM Query \& Procedural Behavior Generation}
\label{sec334}
The plain-language scene description is combined with a predefined internal prompt (i.e., crafted to elicit structured movement instructions) sent via API to the selected LLM, which returns an encoded action plan for each agent. The returned response is formatted using a custom syntax known as the \texttt{SceneDirector} string, which follows the pattern: \texttt{A1\{O1(I, D, S, G, St, B)\}, O2(...)\}, A2\{...\}}, where each entry includes:
\begin{itemize}
    \item \texttt{A}: Agent ID
    \item \texttt{O}: Object ID
    \item \texttt{I}: Interact (T/F)
    \item \texttt{D}: Interaction Duration
    \item \texttt{S}: Movement Speed
    \item \texttt{G}: Grab (T/F)
    \item \texttt{St}: Stationary (T/F)
    \item \texttt{B}: Basic Interaction(T/F) 
\end{itemize}

Based on the above, an example scenario could be: \texttt{A\_1\{Obj\_1(T, 2, 1.5, F, T, F), Obj\_2(F, 1, 1.5, F, F, T)\}, A\_2\{...\}}. In this scenario, agent \texttt{A\_1} interacts with object \texttt{Obj\_1} for two seconds by moving to it at \texttt{1.5x} speed. Since the interaction is stationary, agent \texttt{A\_1} moves to \texttt{Obj\_2} at \texttt{1.5x} speed. Although it is a basic interaction, the agent does not respond and remains at its location for one second. While this is going on, Agent \texttt{A\_2} is performing its interactions in parallel.

\subsubsection{\texttt{SceneDirector} Parsing and Agent Assignment}
\label{sec335}
The \texttt{SceneDirector} component parses the LLM's response, identifies each virtual agent by ID, and assigns each agent a queue of \texttt{destination} structs that define the movement and interaction instructions in order. Each \texttt{destination} contains a reference to the interactable object, flags for interaction, \texttt{grab}, \texttt{basic}, and \texttt{stationary} types, and \texttt{speed} and animation \texttt{duration} for each interaction.

\subsubsection{Movement Execution and Animation Playback}
\label{sec336}
The agent's \texttt{Movement} script executes the \texttt{destination} queue using pathfinding and an animation controller. Each interaction type is handled using dedicated coroutine logic:
\begin{itemize}
    \item \texttt{normal}: Move to object $\rightarrow$ Stop $\rightarrow$ Play animation $\rightarrow$ Wait $\rightarrow$ Move on.
    \item \texttt{grab}: Move to object $\rightarrow$ Grab and attach object $\rightarrow$ Play grab animation while moving to next $\rightarrow$ Drop when done.
    \item \texttt{stationary}: Move to point $\rightarrow$ Stop $\rightarrow$ Play long idle animation (e.g., sleeping, sitting) $\rightarrow$ Wait $\rightarrow$ Continue.
    \item \texttt{basic}: Triggers inverse kinematics-based interaction using the \texttt{InteractionSystem}. Often linked with other scripts (e.g., toggling a light, turning a washing machine on).
\end{itemize}

Each interaction category corresponds to a distinct mode of agent behavior. The \texttt{normal} interaction involves navigating to an object and performing a full-body animation. In contrast, the \texttt{grab} interaction additionally attaches the object, allowing it to be carried to subsequent destinations. The \texttt{stationary} interaction holds the agent at a location for a sustained contextual animation. Finally, the \texttt{basic} interaction triggers short IK-driven contact events that activate object-specific behaviors. All interaction types rely on coroutine-based timing and animation control, allowing them to integrate smoothly with navigation, layered animation blending, and multi-agent sequencing without reintroducing the operational detail described earlier.

When \texttt{grab} and \texttt{stationary} interactions occur concurrently, the animation system resolves them through layered blending; \texttt{grab} animations are assigned to the upper-body layer (Layer 1), while \texttt{stationary} actions are maintained on the lower-body layer (Layer 0), allowing both to play simultaneously without conflict. Timing is coordinated so that the carried object remains attached until the \texttt{grab} sequence fully completes, ensuring consistent visual behavior even when animation durations overlap.

\subsubsection{Finalization}
\label{sec337}
\begin{figure*}[!htb]
    \centering
    \begin{subfigure}[b]{.33\textwidth}
        \includegraphics[width=\textwidth]{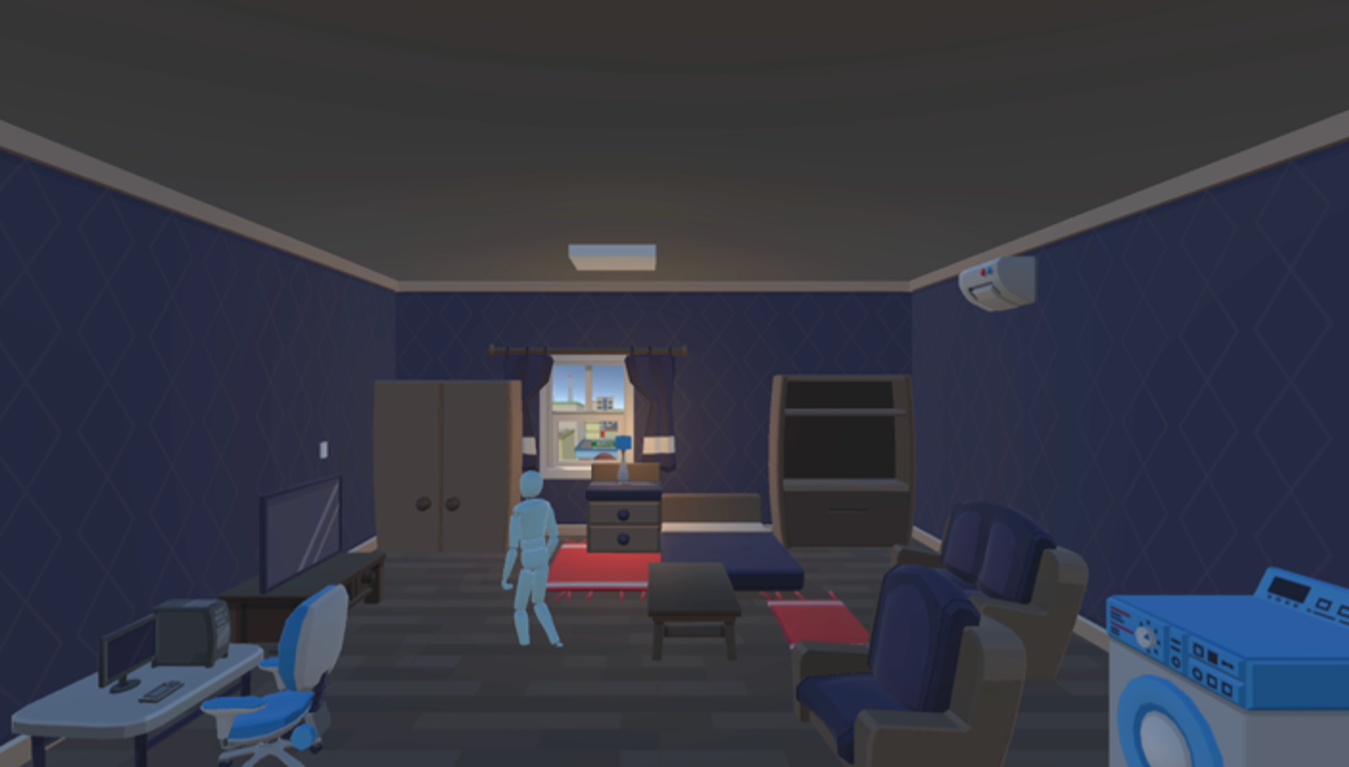}
        \caption{}
        \label{fig:gull}
    \end{subfigure}
    \begin{subfigure}[b]{.33\textwidth}
        \includegraphics[width=\textwidth]{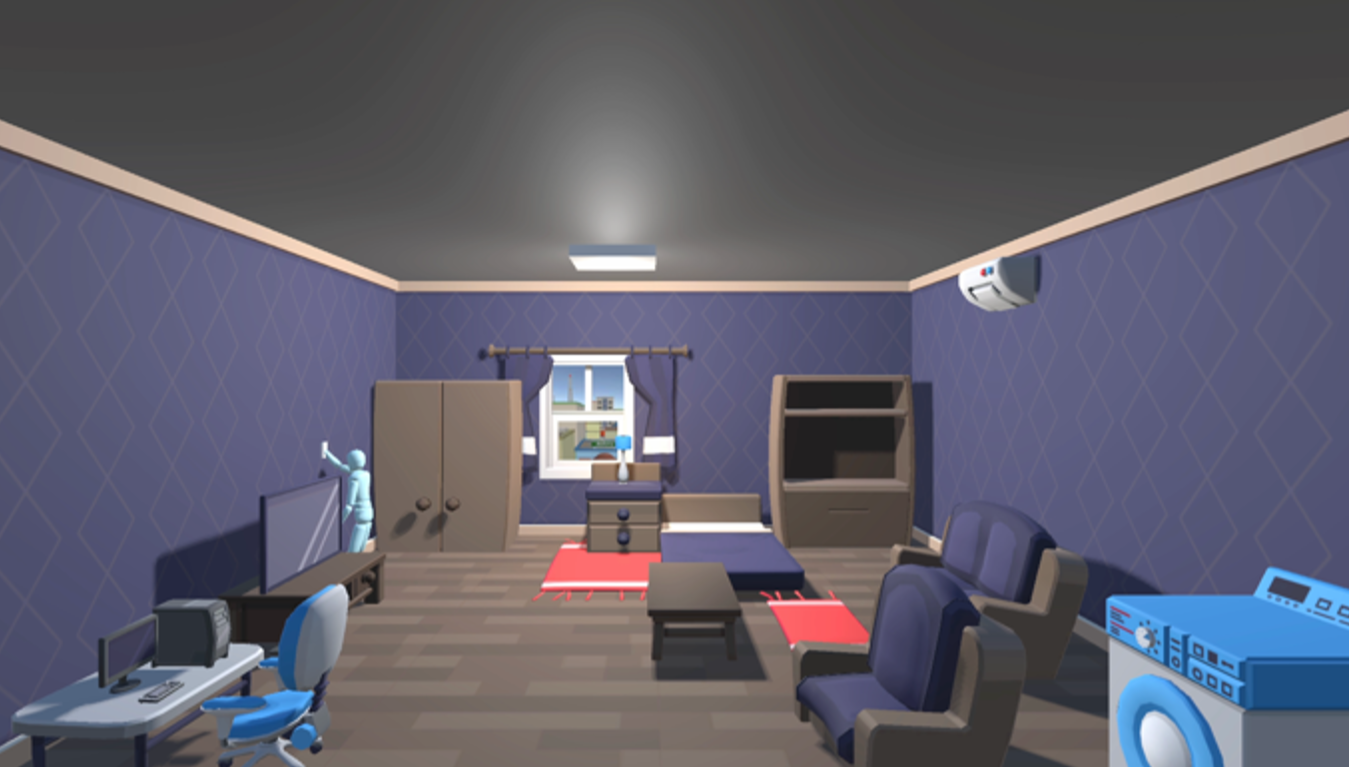}
        \caption{}
        \label{fig:tiger}
    \end{subfigure}
    \begin{subfigure}[b]{.33\textwidth}
        \includegraphics[width=\textwidth]{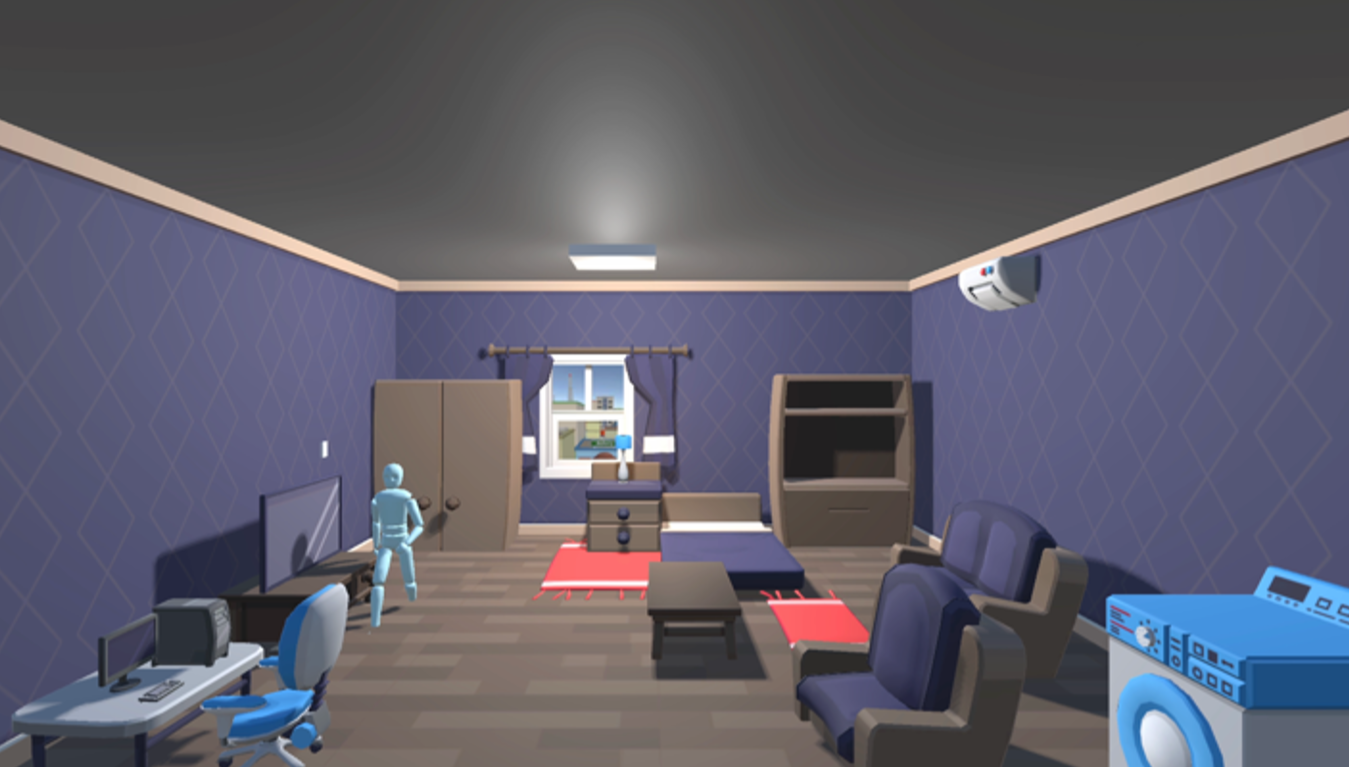}
        \caption{}
        \label{fig:mouse}
    \end{subfigure}
     \begin{subfigure}[b]{.33\textwidth}
        \includegraphics[width=\textwidth]{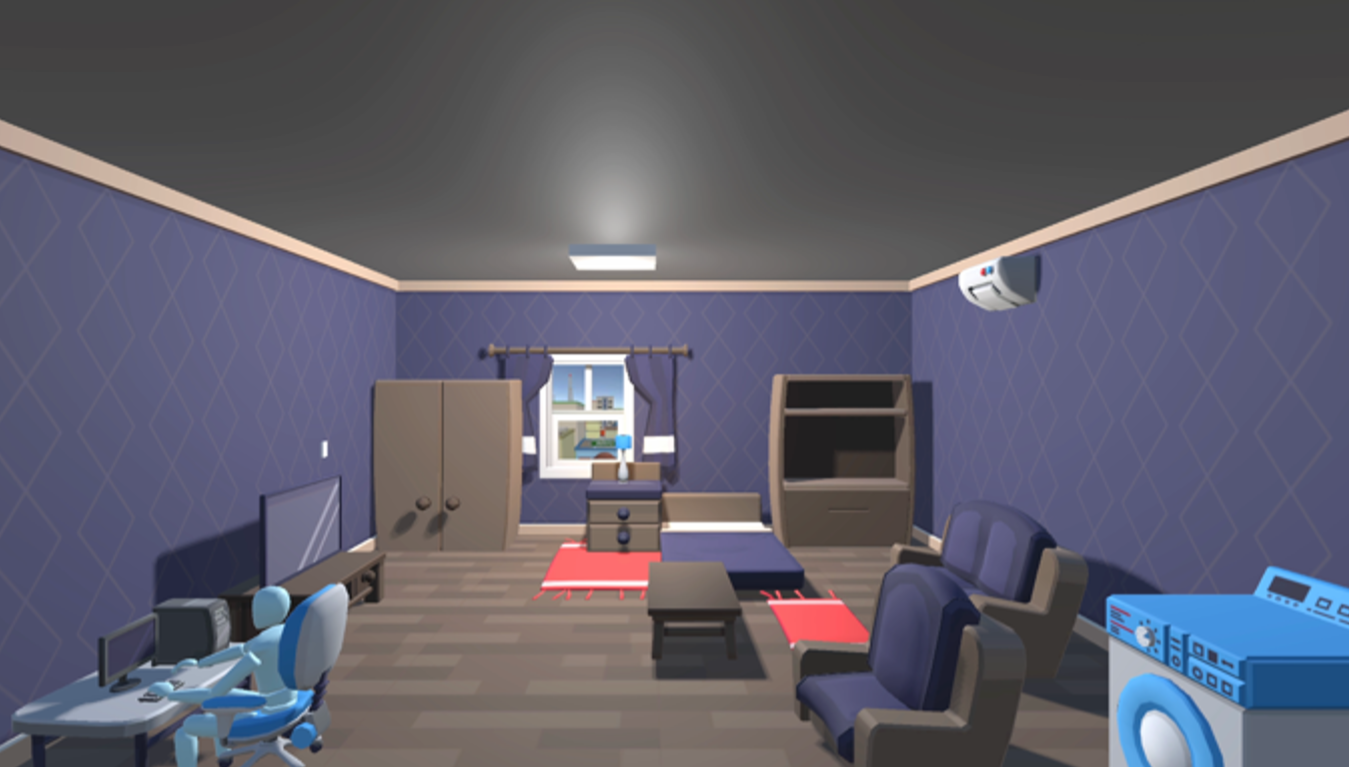}
        \caption{}
        \label{fig:mouse}
    \end{subfigure}
     \begin{subfigure}[b]{.33\textwidth}
        \includegraphics[width=\textwidth]{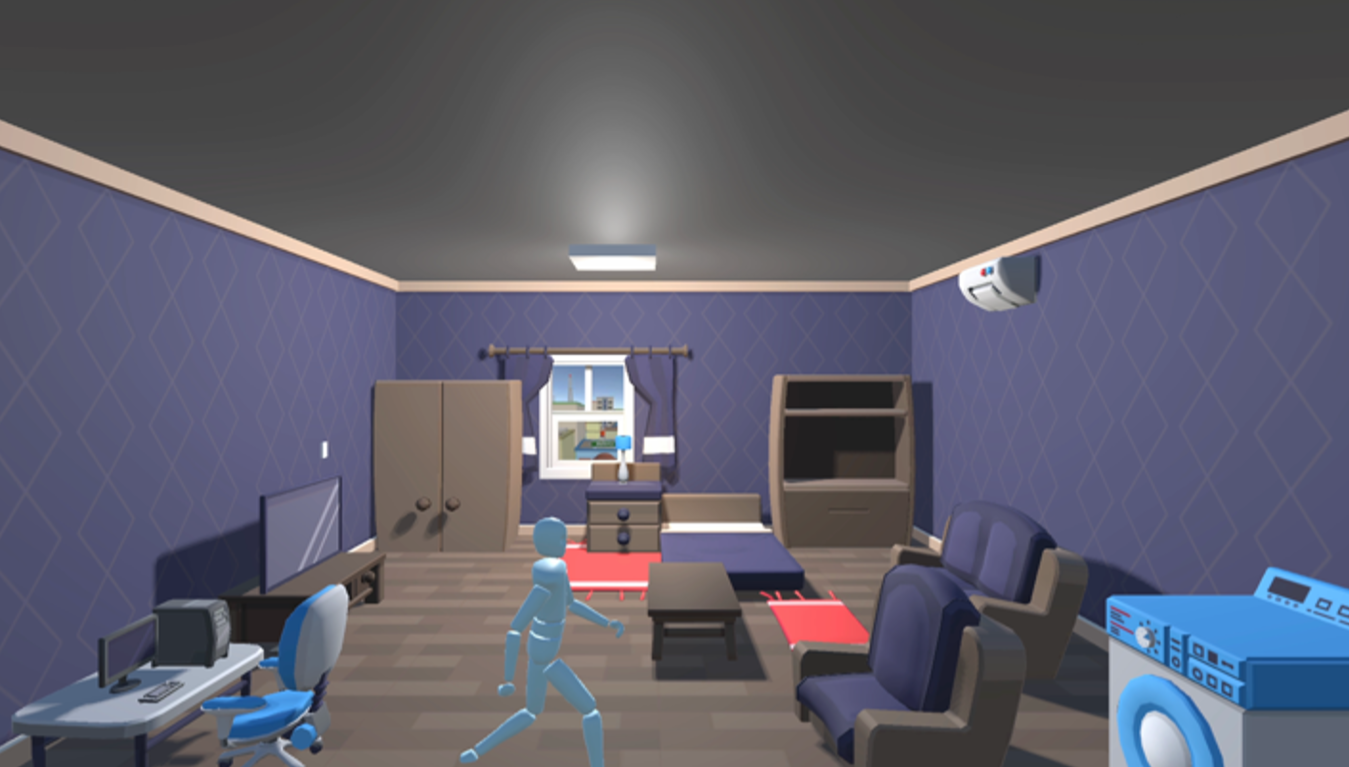}
        \caption{}
        \label{fig:mouse}
    \end{subfigure}
     \begin{subfigure}[b]{.33\textwidth}
        \includegraphics[width=\textwidth]{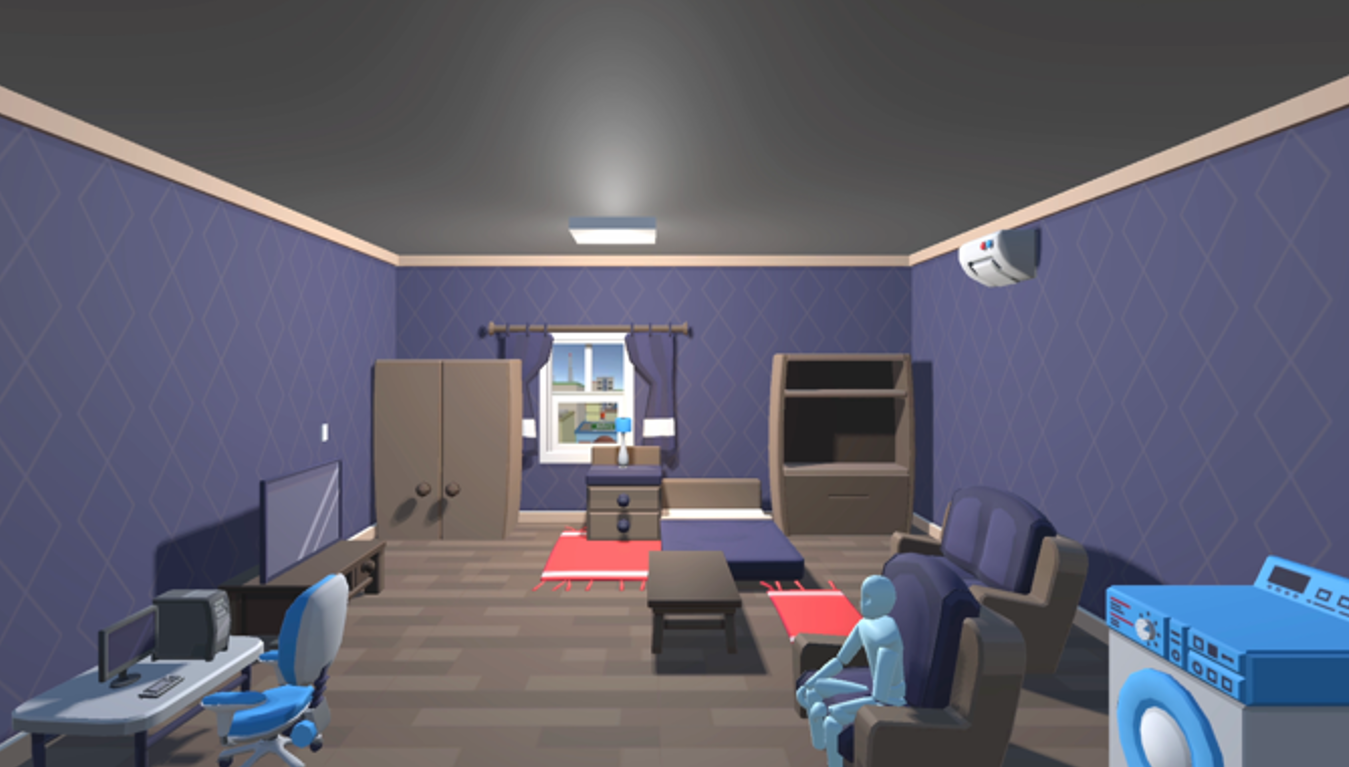}
        \caption{}
        \label{fig:mouse}
    \end{subfigure}
    \caption{Example generated output based on metadata description.}
    \label{fig5}
\end{figure*}

Once all interactions for a character are completed, the character returns to an idle or default state, the pathfinding is stopped, and animations are reset to \texttt{locomotion}. An example scenario is shown in Figure~\ref{fig5}. This example illustrates the flow of the final output. The agent moves to the light switch (a), the agent toggles the light switch (b), the agent moves to the next destination (c), the agent sits down and uses the computer (d), the agent moves to the next destination (e), and the agent sits down on the chair (f). While this is generated for a single agent, the same processes occur for all agents in the scene when multiple agents are present.

\subsection{Implementation Details }
\label{sec34}
The system was implemented in the Unity game engine version 2022.3.9, which served as the primary environment for scene construction, runtime execution, and agent-object interaction management. All system logic was implemented in \texttt{C\#} using JetBrains Rider as the main development environment. Four lightweight LLMs, ChatGPT (GPT-4.1 mini) from OpenAI, Claude (Claude Sonnet 4.5) from Anthropic, Gemini (Gemini 2.5 Flash) from Google, and Grok (Grok 4.1 Fast) from xAI, were integrated through their respective APIs to convert serialized scene descriptions into structured interaction logic. The animation system incorporated an IK system for IK-based adjustments, enabling precise pose alignment during \texttt{basic} interactions. Scene construction was supported by grid-aligned placement of agents and objects. Animation assets were sourced from Mixamo, supplying humanoid clips for idle, interaction, and grab behaviors. \texttt{ScriptableObjects} were used to store metadata for each prefab, ensuring scalable, semantically consistent configuration of agents and interactable objects throughout the scene.

\subsection{Estimating the Number of Possible Scenarios}
\label{sec35}
To quantify the diversity of possible outcomes in this procedural generation system, we analyzed the factors contributing to scenario variation. Unlike traditional branching logic systems, our approach allows for dynamic combinations of agents, objects, and interaction contexts. Each object in the scene supports only one type of interaction (e.g., \texttt{grab}, \texttt{stationary}, or \texttt{basic}); the richness of the system arises from how these interactions are sequenced, layered, and distributed across agents. Interactions can vary in duration, occur at different locations, and may overlap when compatible (e.g., \texttt{grab} or \texttt{stationary} interactions). Additionally, the LLM-based story planner considers the contextual positioning of each object and agent to compose semantically coherent sequences. Considering these variables, we derive the following formula to estimate the number of potential interaction scenarios:
\begin{equation}
Total Scenarios = (m \cdot v \cdot d)^n
\end{equation}
where $m$ denotes the number of interactable objects, $v$ denotes the number of spatial placements or contextual variants per object, $d$ denotes the number of distinct duration or timing variations for interactions, and $n$ denotes the number of virtual agents in the scene. This formulation captures not just object-agent assignments but also spatial arrangement, interaction timing, and action sequencing, all of which contribute to the generation of emergent narratives through LLM-driven agent-based narratives planning.

\section{Evaluation and Results}
\label{sec4}
The evaluation focuses on benchmarking the performance of four LLMs, ChatGPT (gpt-4.1-mini), Claude (claude-sonnet-4-5), Gemini (gemini-2.5-flash), and Grok (grok-4-1-fast) when integrated into the procedural agent-based activity generation system. Each model was tested across a fixed set of scene-complexity scenarios to measure response latency, consistency, and validity of output. The five controlled test scenarios were designed to represent increasing narrative and computational complexity:
\begin{itemize}
    \item \textbf{1O-1A:} 1 Object and 1 Agent
    \item \textbf{5O-1A:} 5 Objects and 1 Agent
    \item \textbf{5O-2A:} 5 Objects and 2 Agents
    \item \textbf{5O-5A:} 5 Objects and 5 Agents
    \item \textbf{10O-5A:} 10 Objects and 5 Agents
\end{itemize}

\begin{figure*}[!htb]
    \centering
    \begin{subfigure}[b]{.33\textwidth}
        \includegraphics[width=\textwidth]{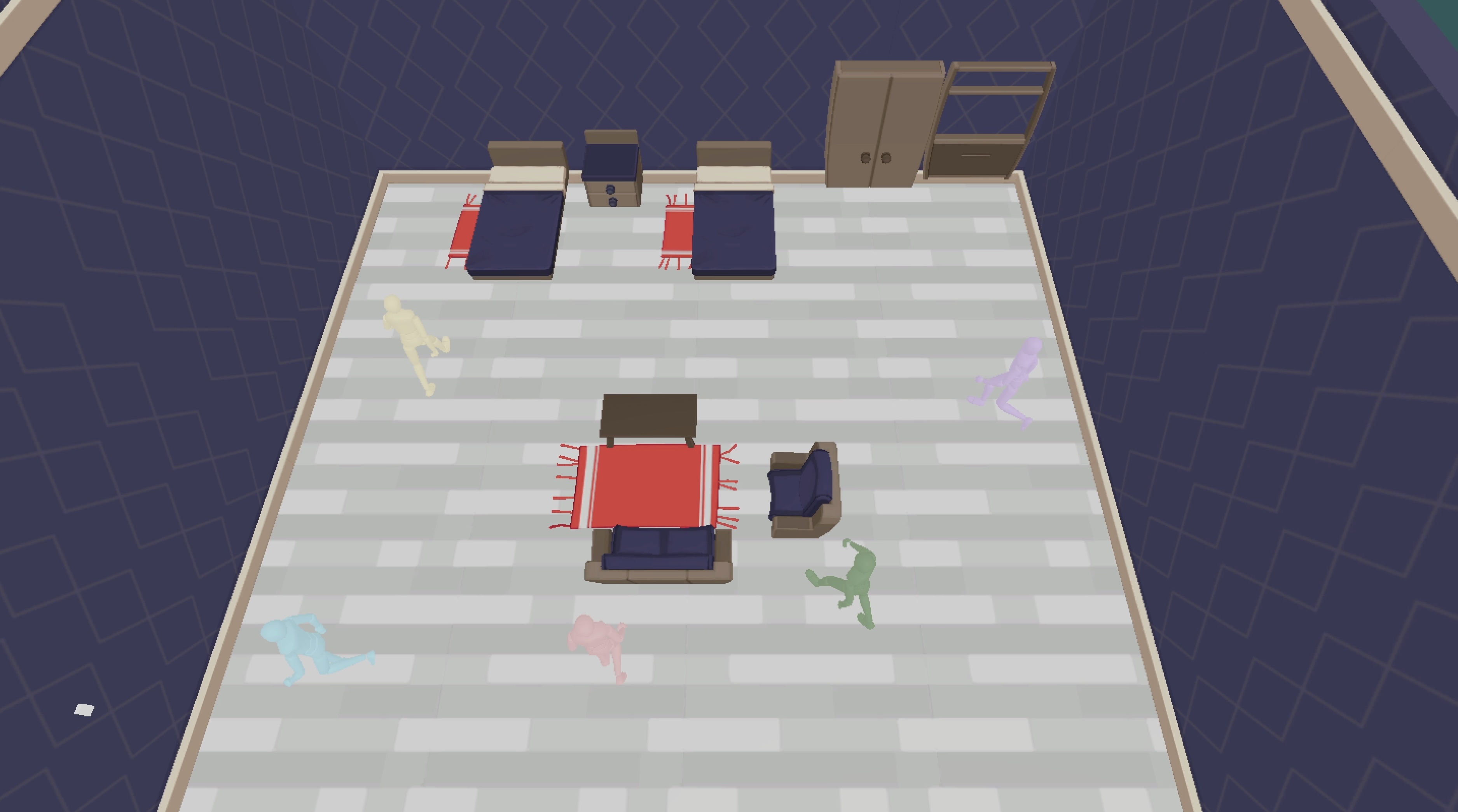}
        \caption{}
        \label{fig:gull}
    \end{subfigure}
    \begin{subfigure}[b]{.33\textwidth}
        \includegraphics[width=\textwidth]{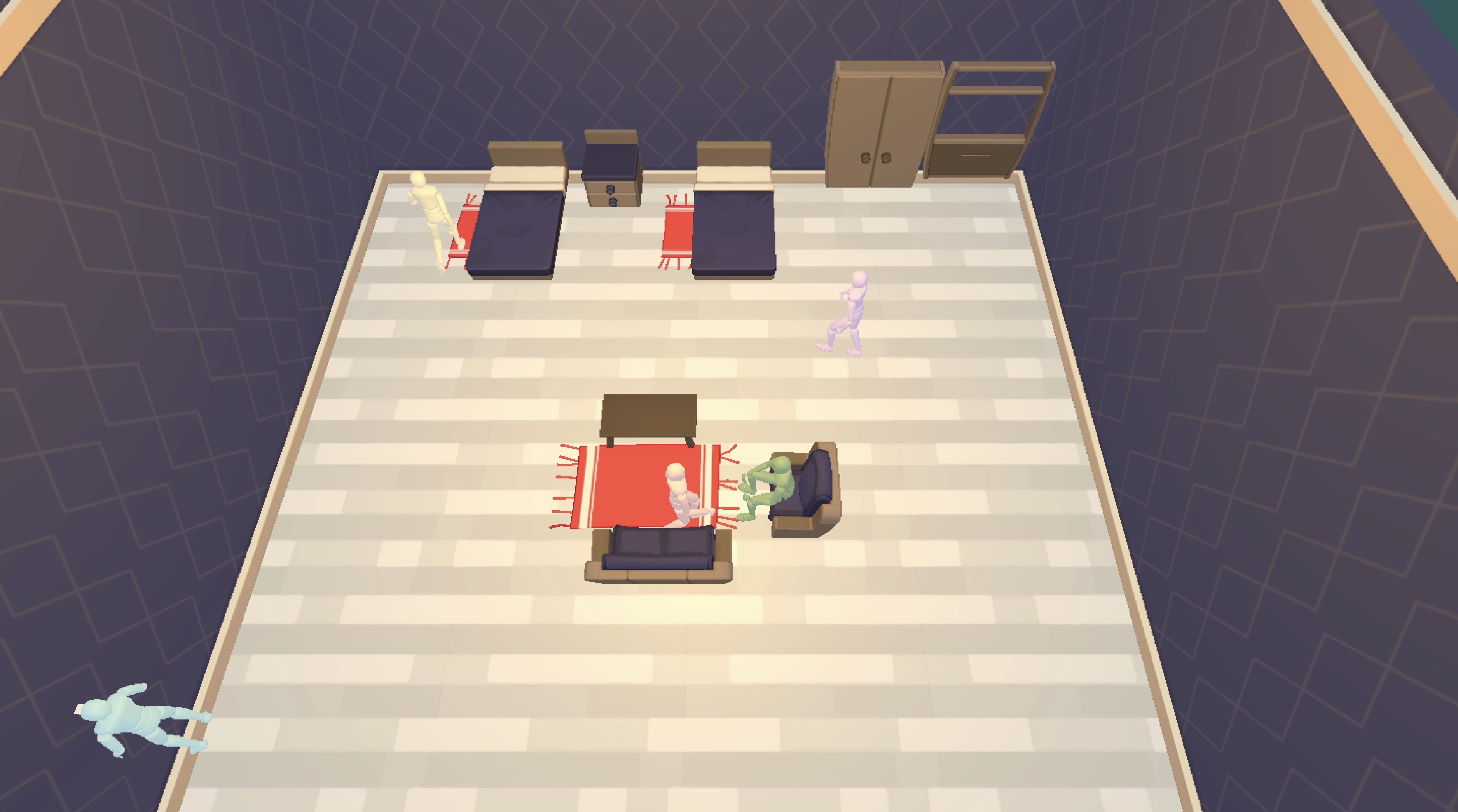}
        \caption{}
        \label{fig:tiger}
    \end{subfigure}
    \begin{subfigure}[b]{.33\textwidth}
        \includegraphics[width=\textwidth]{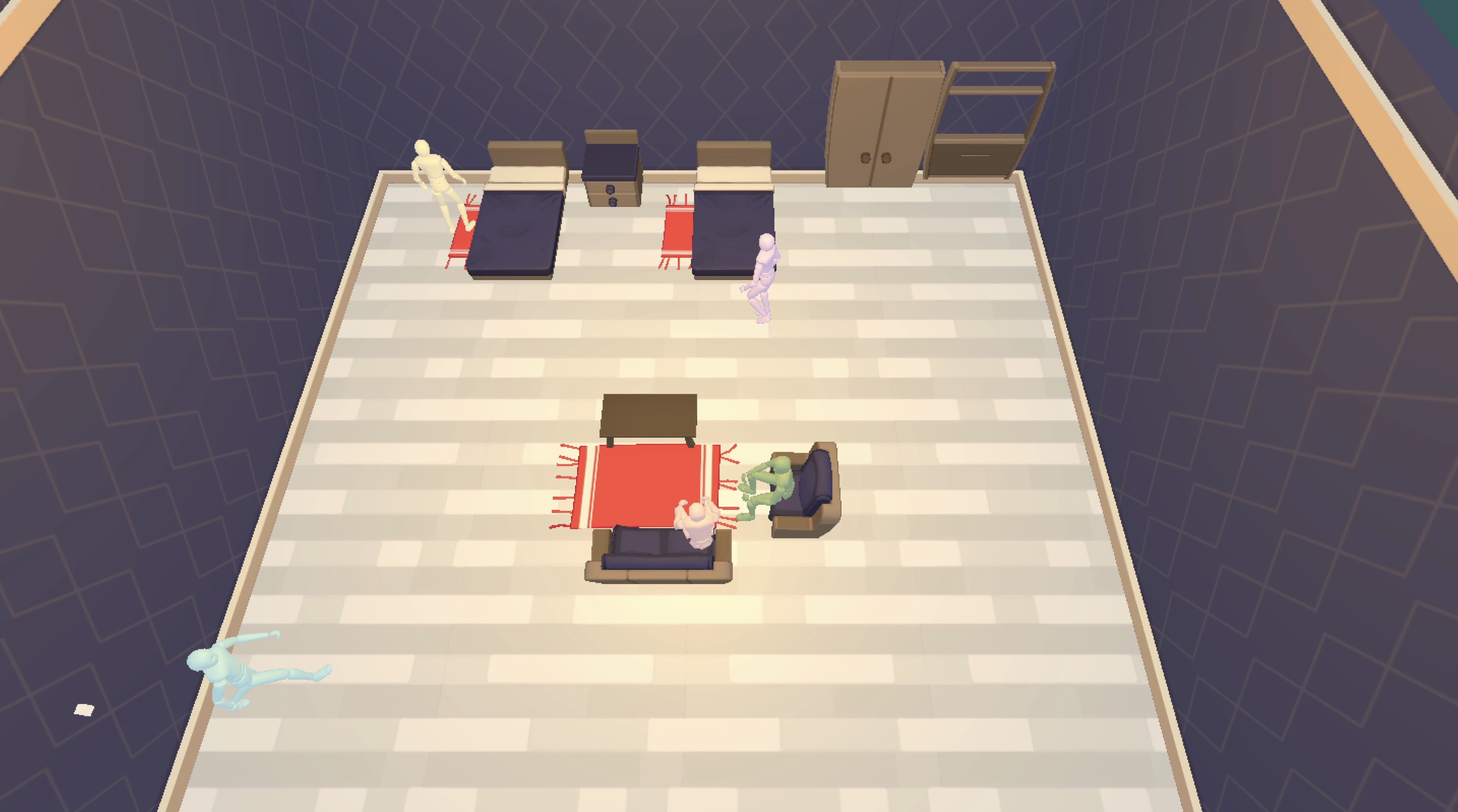}
        \caption{}
        \label{fig:mouse}
    \end{subfigure}
     \begin{subfigure}[b]{.33\textwidth}
        \includegraphics[width=\textwidth]{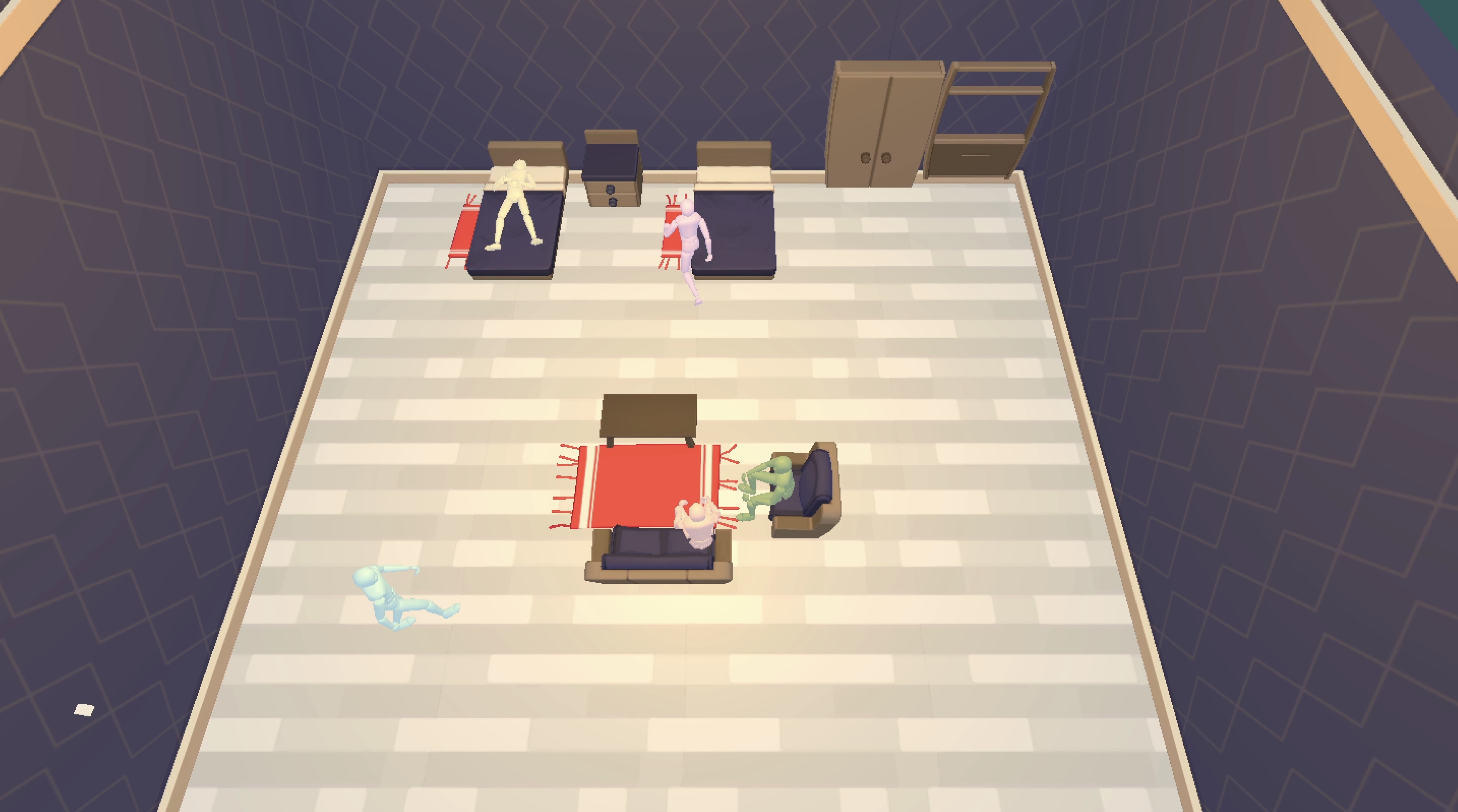}
        \caption{}
        \label{fig:mouse}
    \end{subfigure}
     \begin{subfigure}[b]{.33\textwidth}
        \includegraphics[width=\textwidth]{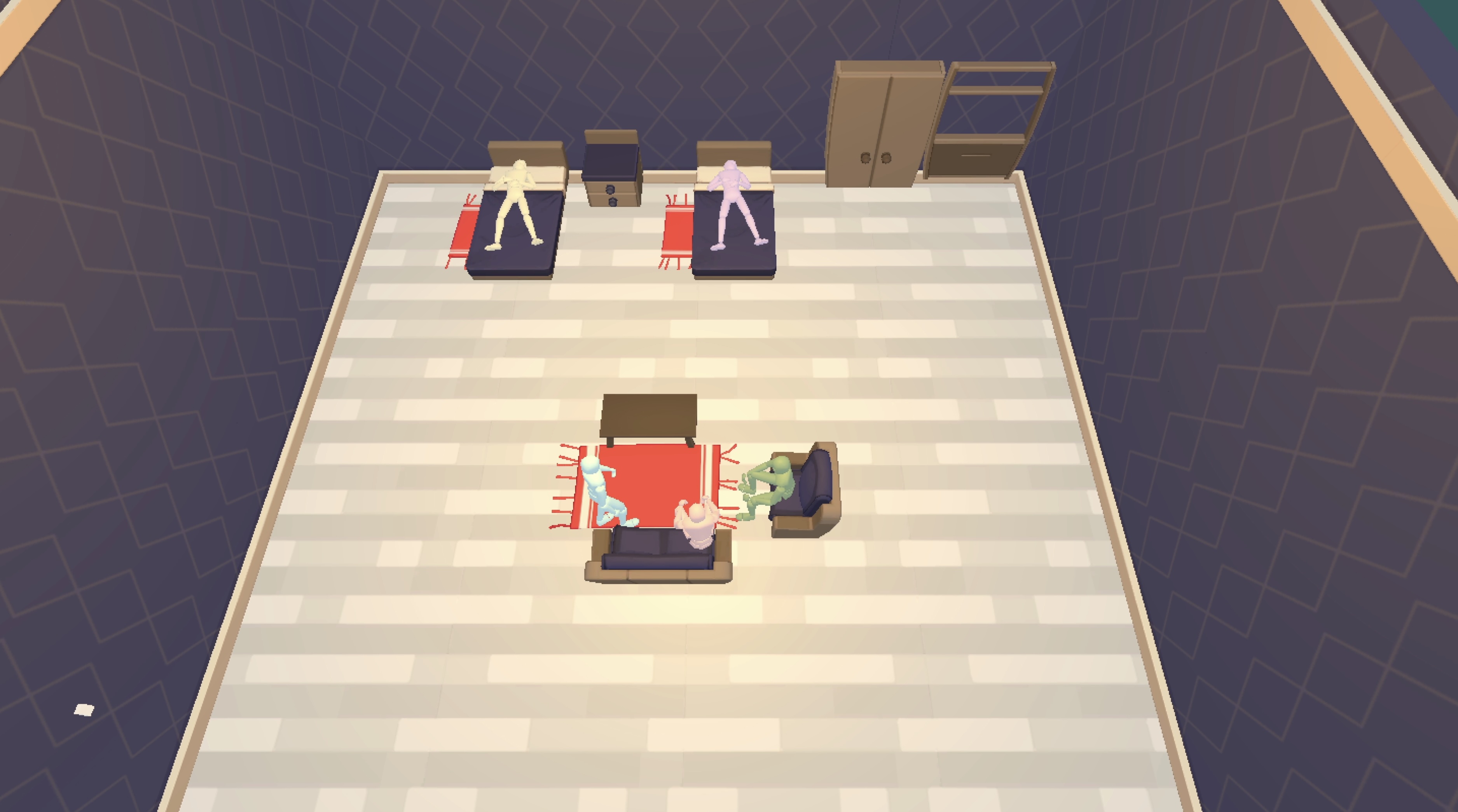}
        \caption{}
        \label{fig:mouse}
    \end{subfigure}
     \begin{subfigure}[b]{.33\textwidth}
        \includegraphics[width=\textwidth]{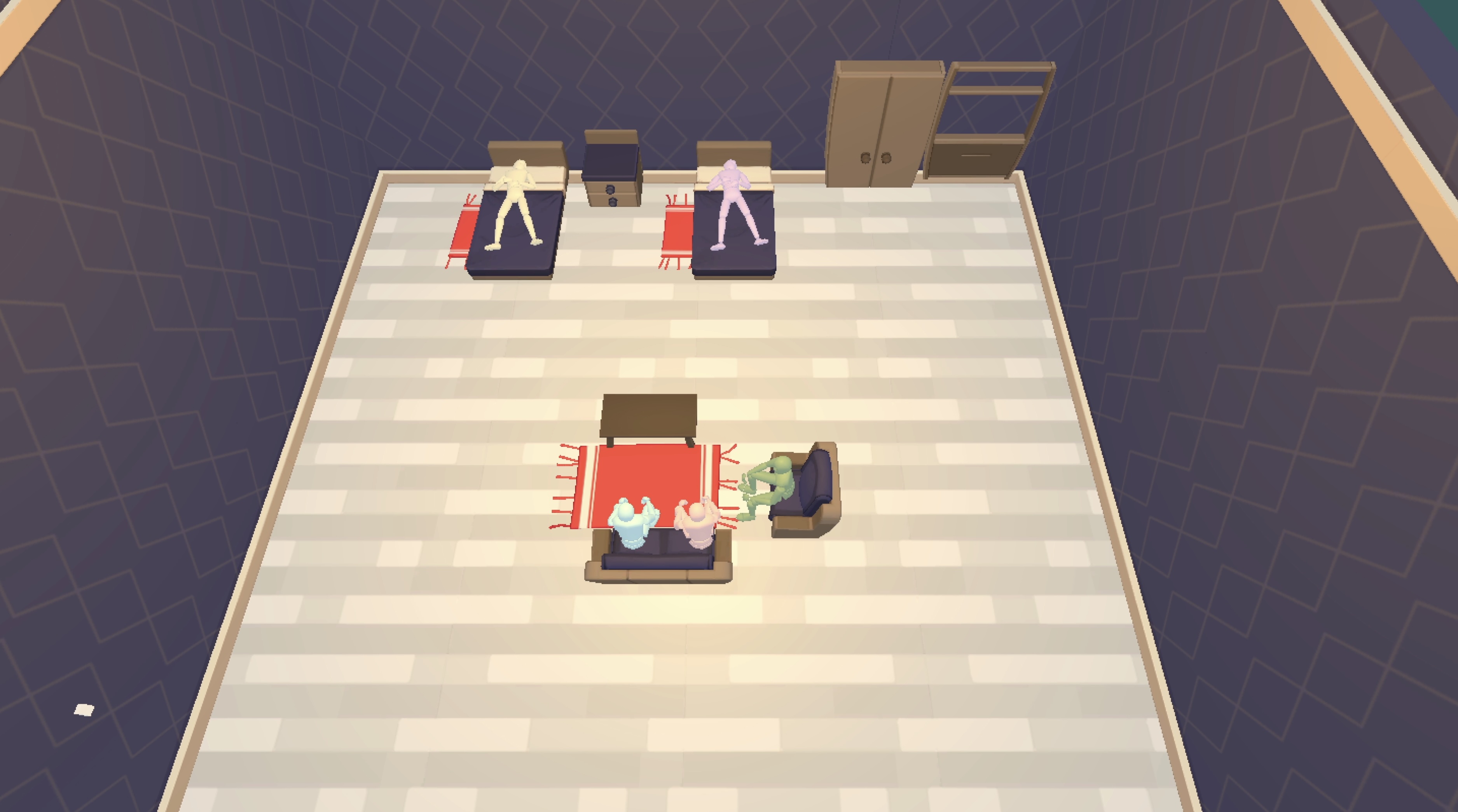}
        \caption{}
        \label{fig:mouse}
    \end{subfigure}
    \caption{Generated output for the \textbf{5O-5A} scenario.}
    \label{fig8}
\end{figure*}

Each configuration was converted into a structured natural-language prompt describing spatial relationships, object affordances, and agent roles. All models received identical prompt text to ensure evaluation consistency. An example of a generated 
\textbf{5O-5A} multi-agent scene is shown in Figure~\ref{fig8}, and an example of a generated \textbf{10O-5A} multi-agent scene is shown in Figure~\ref{fig7}. Moreover, examples of generated output are provided in the supplementary video.

\begin{figure*}[!htb]
    \centering
    \begin{subfigure}[b]{.33\textwidth}
        \includegraphics[width=\textwidth]{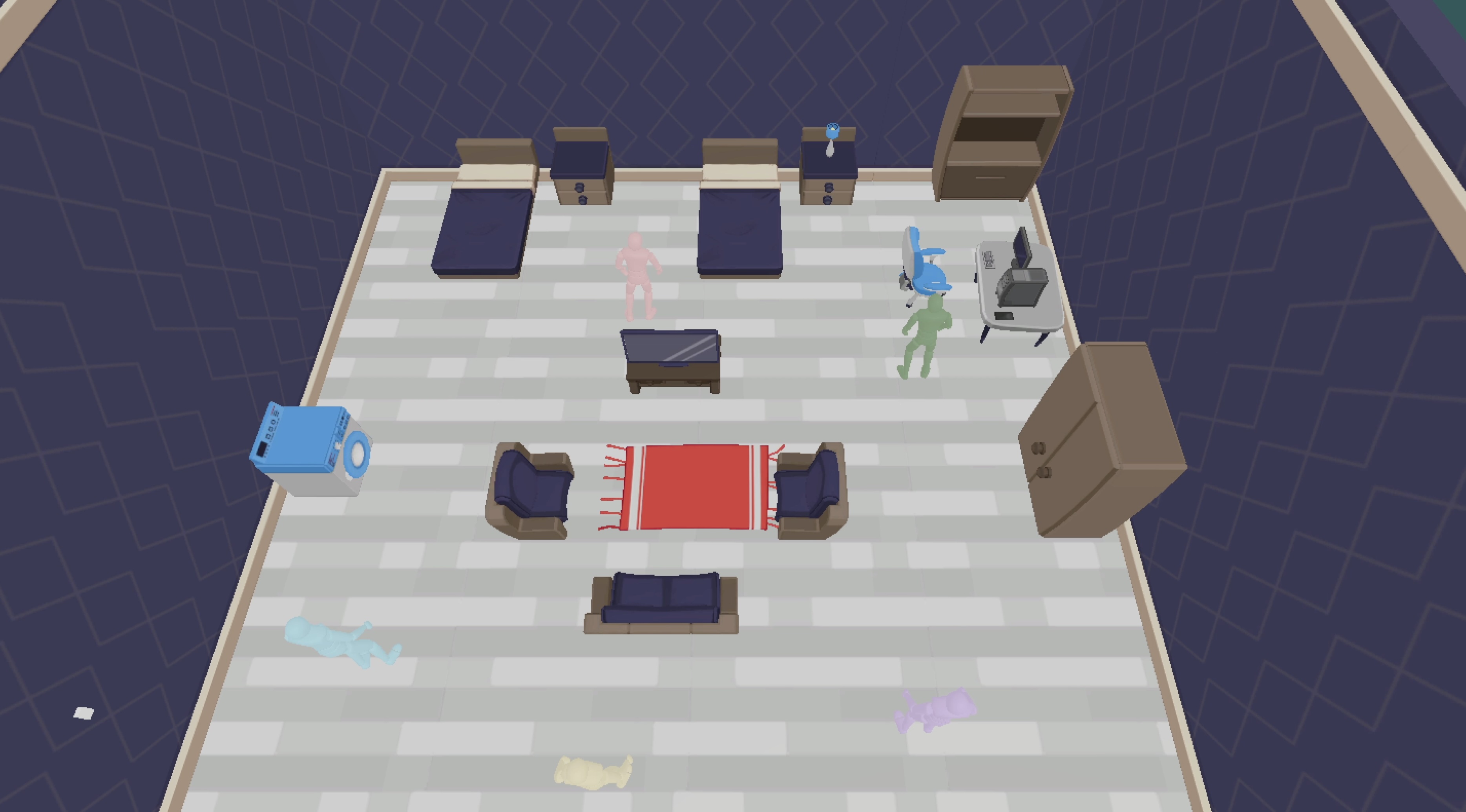}
        \caption{}
        \label{fig:gull}
    \end{subfigure}
    \begin{subfigure}[b]{.33\textwidth}
        \includegraphics[width=\textwidth]{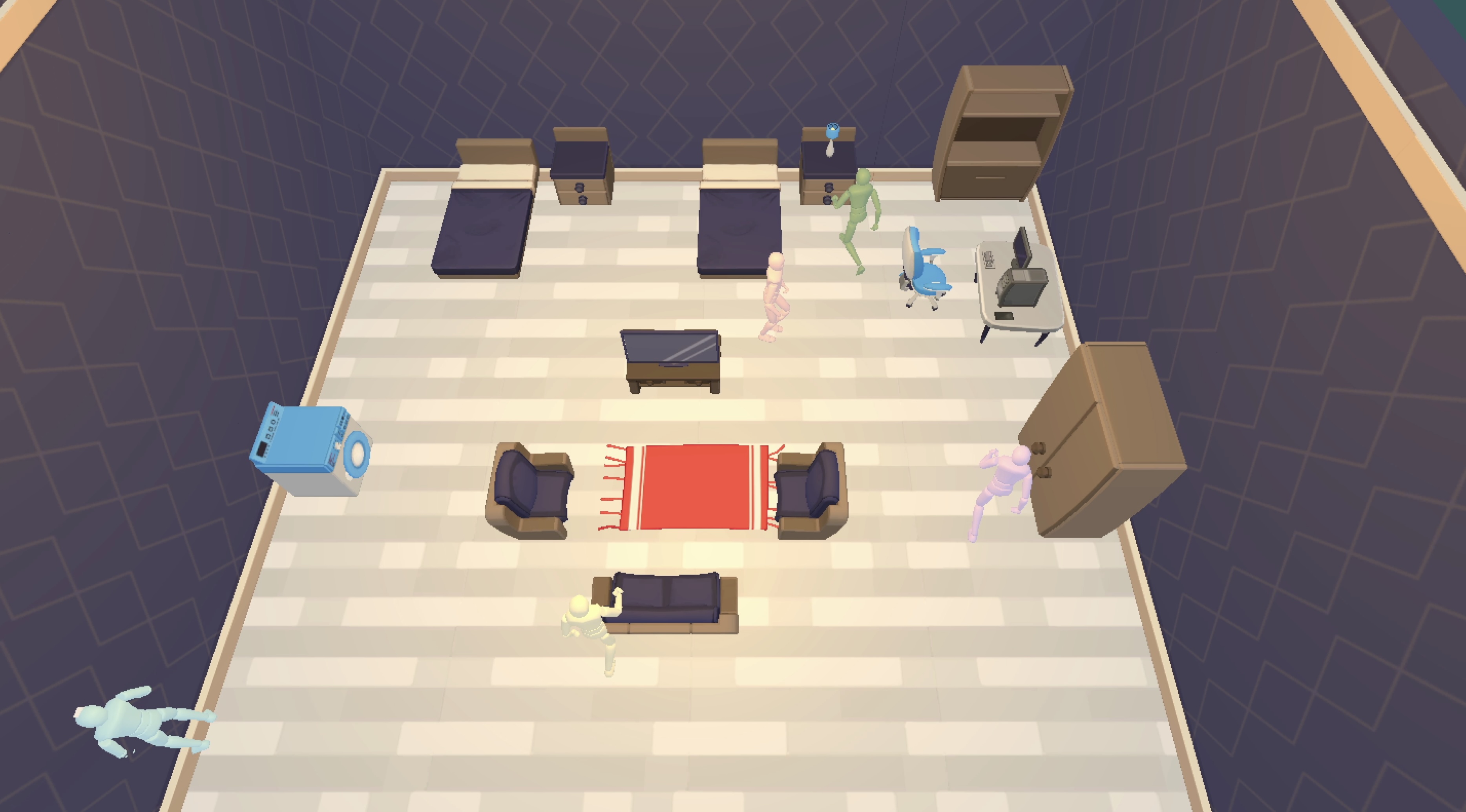}
        \caption{}
        \label{fig:tiger}
    \end{subfigure}
    \begin{subfigure}[b]{.33\textwidth}
        \includegraphics[width=\textwidth]{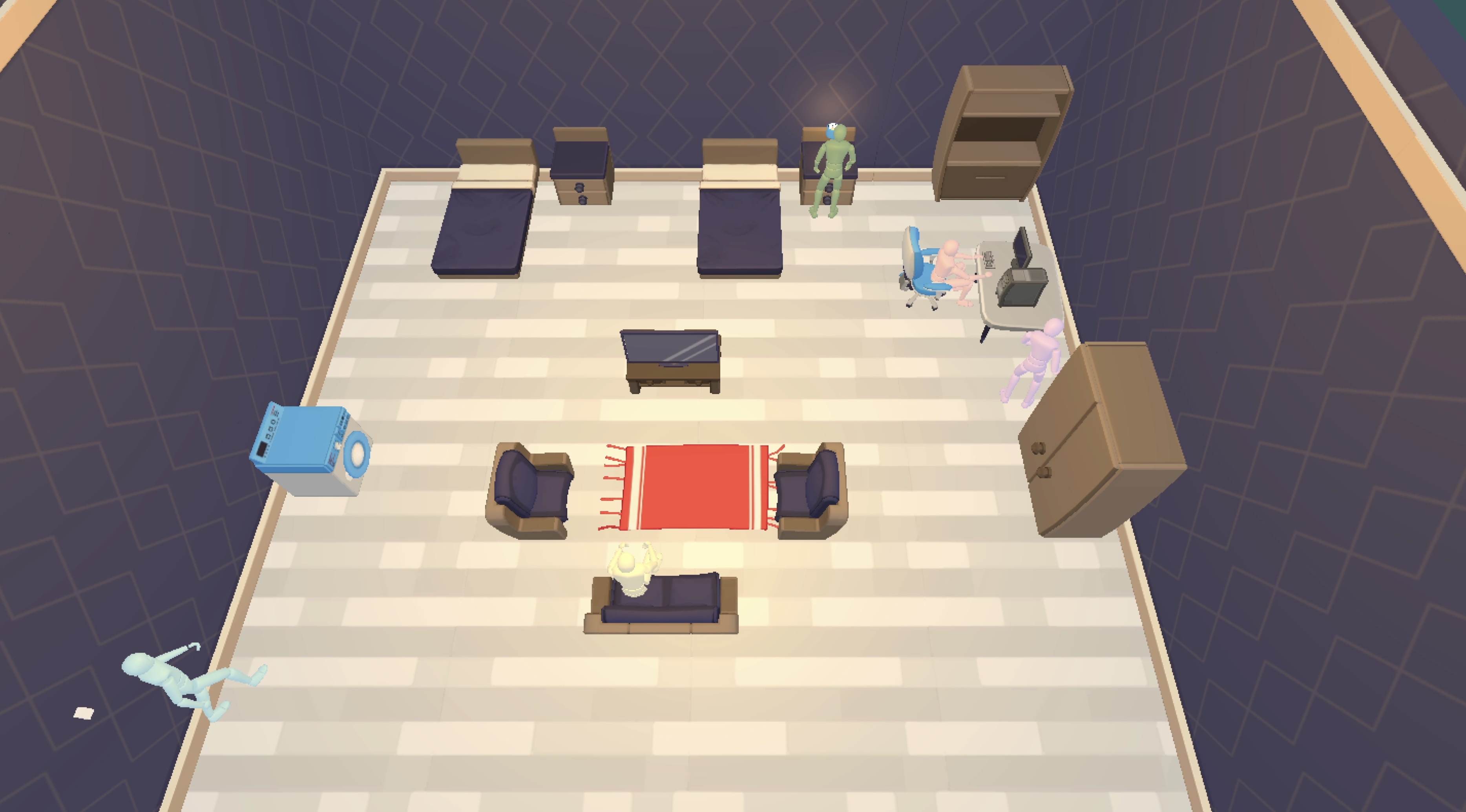}
        \caption{}
        \label{fig:mouse}
    \end{subfigure}
     \begin{subfigure}[b]{.33\textwidth}
        \includegraphics[width=\textwidth]{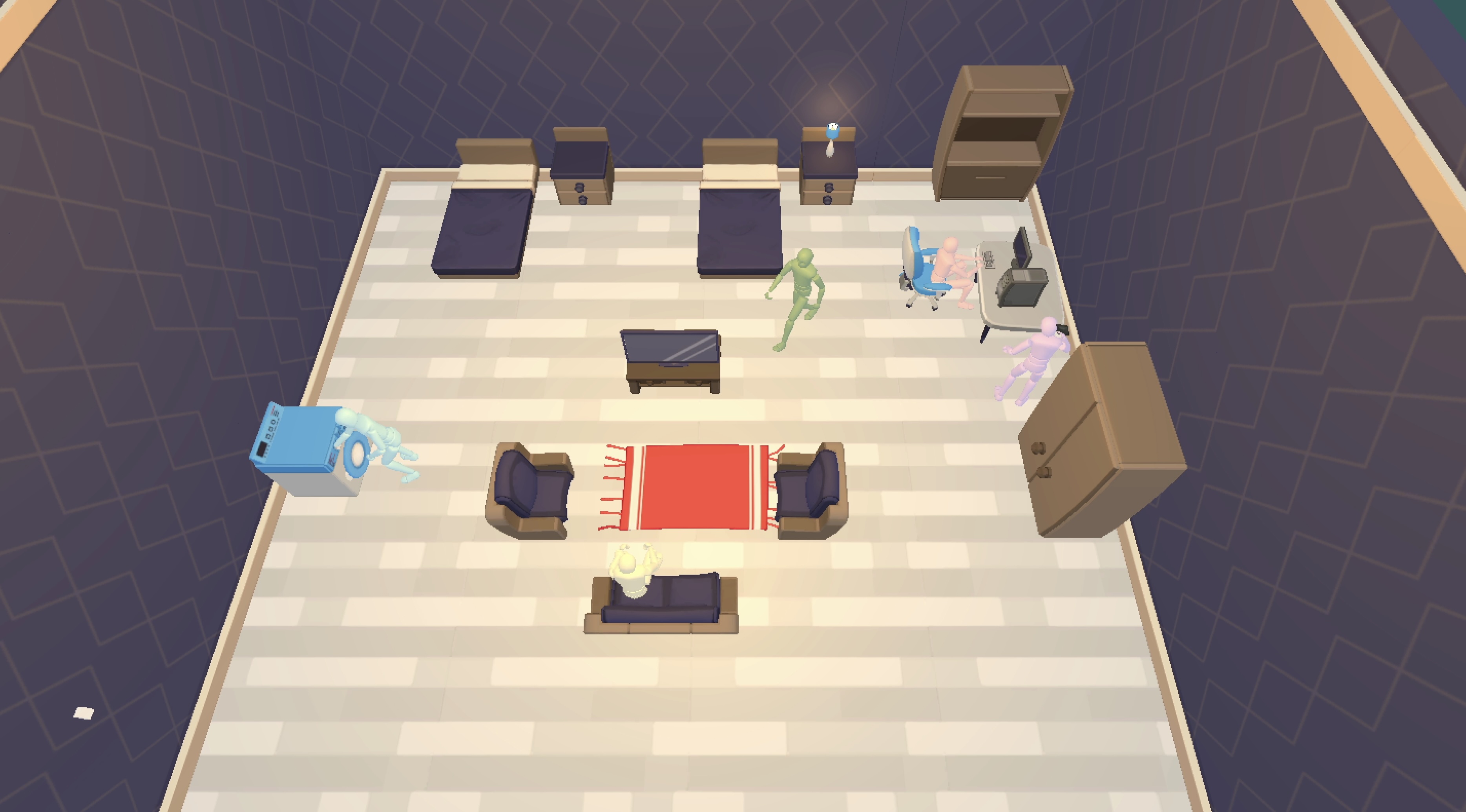}
        \caption{}
        \label{fig:mouse}
    \end{subfigure}
     \begin{subfigure}[b]{.33\textwidth}
        \includegraphics[width=\textwidth]{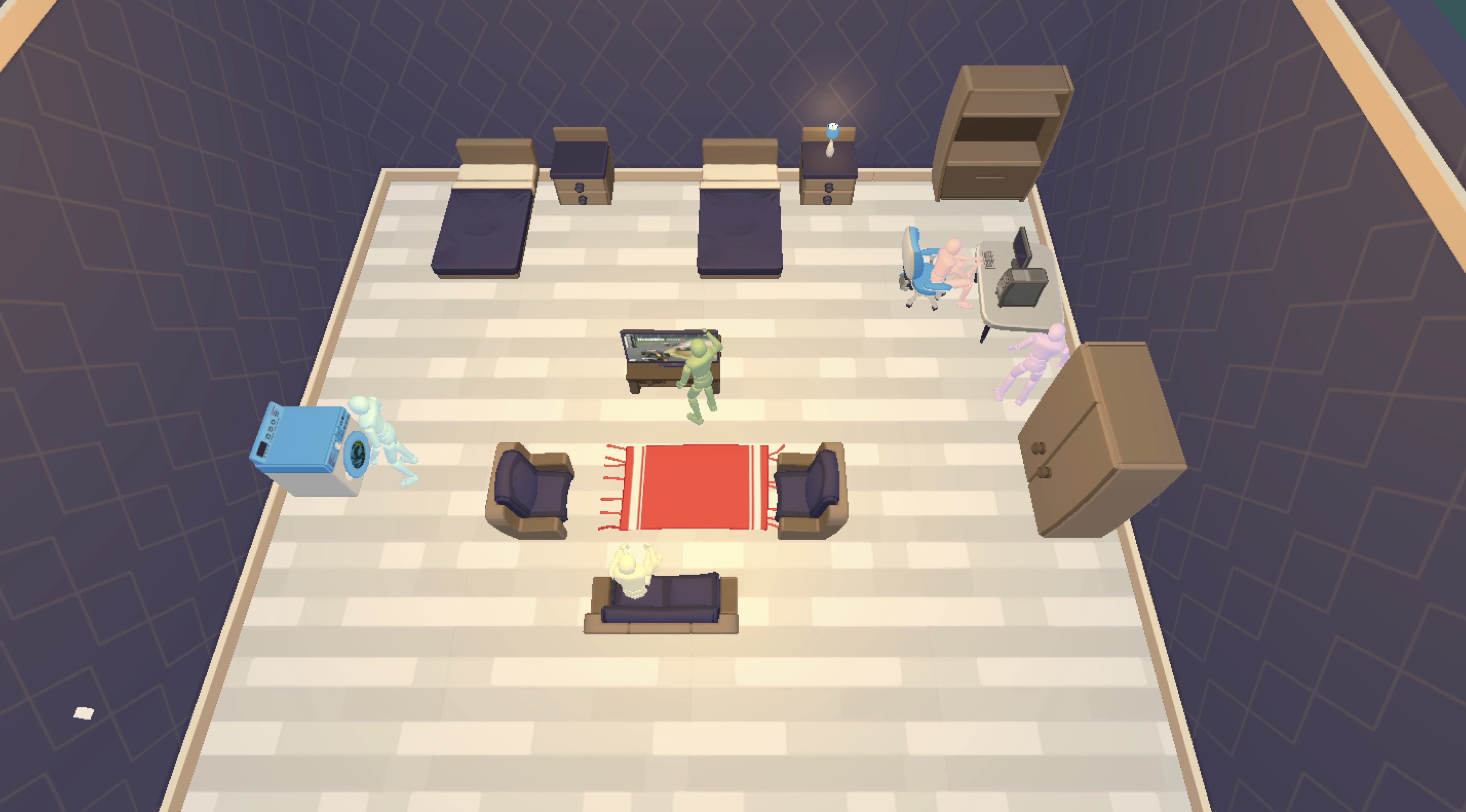}
        \caption{}
        \label{fig:mouse}
    \end{subfigure}
     \begin{subfigure}[b]{.33\textwidth}
        \includegraphics[width=\textwidth]{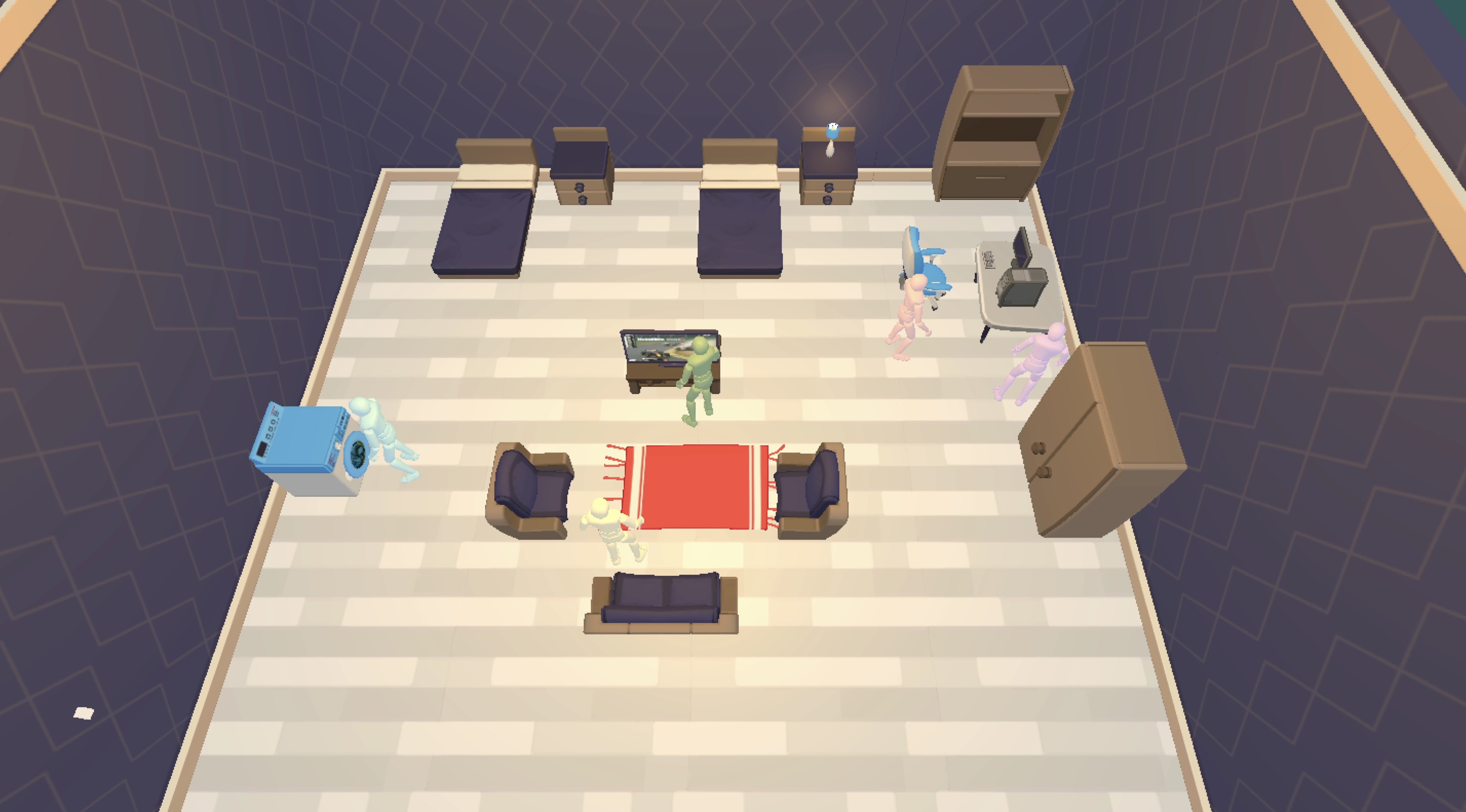}
        \caption{}
        \label{fig:mouse}
    \end{subfigure}
    \caption{Generated output for the \textbf{10O-5A} scenario.}
    \label{fig7}
\end{figure*}

\subsection{Testing Procedure}
\label{sec41}
For each test scenario, an identical natural language prompt was submitted to all four language models in sequence via the automated testing interface. The system measured the response time from the initiation of the request (since the processing internally is done by the LLM, which is not accessible to the API directly, we are timing the combined end-to-end response time, from sending the request to receiving the response), LLM processing, to the receipt of a complete output, and recorded it in seconds. To mitigate the influence of transient network variability, each model was tested five times per scenario, and the average was used in the final analysis.

In addition to timing measurements, each response was examined for structural validity, ensuring that the generated output adhered to the expected formatting required by the Unity parser. Responses that met these criteria were further reviewed for adherence to basic instruction and for the coherence of the resulting action sequence, allowing the evaluation to assess whether increasing scene complexity affected the models' ability to produce logically consistent and syntactically usable behavior plans. Only responses satisfying these validity requirements were retained in the final dataset.

\subsection{Data Analysis}
\label{sec42}
Following data collection, the recorded response times were organized into a comparative dataset structured by model and scenario. For each model-scenario pair, the five timing samples were used to compute descriptive statistics, including the mean response time (i.e., processing and response), enabling a quantitative assessment of model efficiency and stability. Scenario-level timing patterns were examined by comparing man values. In parallel with the timing analysis, the retained outputs were reviewed qualitatively to confirm that they were structurally valid, consistent with the prompt constraints, and fully parsable by the \texttt{SceneDirector}. These quantitative and observational evaluations served as the basis for the performance comparison.

\subsection{Results}
\label{sec43}
The evaluation produced quantitative timing measurements and qualitative assessments of output validity across all scenarios. Here, we present the results of the model comparisons, with each scenario summarized in terms of the correctness of the generated \texttt{SceneDirector} strings and the processing time and its variability.

\subsubsection{Correctness of the Generated \texttt{SceneDirector} Strings}
\label{sec41}
Across all evaluated scenarios, all four models consistently produced \texttt{SceneDirector} outputs that were structurally valid, semantically coherent, compliant with prompt constraints, and fully parsable by the developed system, with no responses discarded for structural or semantic reasons. We provide example outputs for each examined scenario in the supplementary materials document. In the simpler \textbf{1O-1A} and \textbf{5O-1A} settings, models generated coherent interaction sequences aligned with the scene descriptions despite increasing object density. This robustness extended to multi-agent configurations: in both the \textbf{5O-2A} and \textbf{5O-5A} scenarios, models successfully coordinated multiple agents, adhered to the Unity parser's formatting requirements, and produced coherent agent-based interaction plans consistent with the textual scene descriptions. Even in the most complex scenario, \textbf{10O-5A}, featuring five agents interacting with ten objects, all models maintained structural correctness, narrative coherence, and adherence to interaction rules, resulting in fully executable \texttt{SceneDirector} action sequences.

\subsubsection{Processing Time}
\label{sec42}
\begin{figure}[!htb]
 \centering 
 \includegraphics[width=\columnwidth]{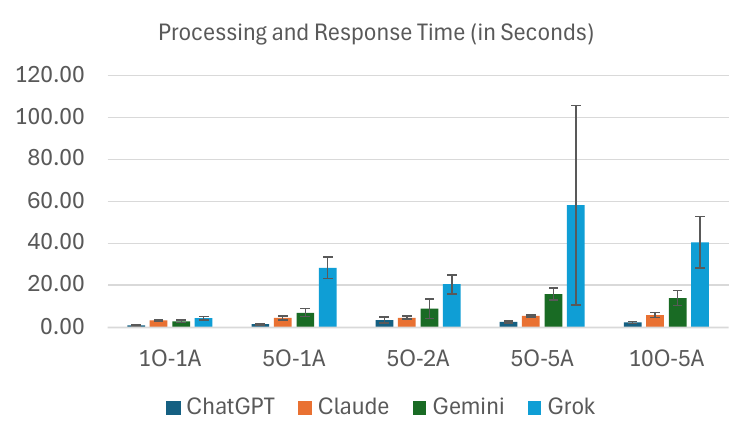}
 \caption{Bar chart illustrating the obtained performance of the four examined LLMs across the five scenarios.}
 \label{fig6}
\end{figure}

We computed the processing time for the four examined LLMs across five runs for each scenario. The results are summarized in Table~\ref{fig3} and Figure~\ref{fig6}.

\begin{table*}[!htb]
\caption{Processing and response time (in seconds) of each examined scenario across the examined LLM providers.}
\label{tab3}
\centering
\begin{tabular}{lrrrrrrrr}
\toprule
       & \multicolumn{2}{c}{\textbf{ChatGPT}} & \multicolumn{2}{c}{\textbf{Claude}} & \multicolumn{2}{c}{\textbf{Gemini}} & \multicolumn{2}{c}{\textbf{Grok}} \\
\midrule
& $M$            & $SD$           & $M$            & $SD$          & $M$            & $SD$          & $M$           & $SD$         \\
\midrule
\textbf{1O-1A}  & .79          & .13          & 3.27         & .45         & 2.94         & .71         & 4.38        & .79        \\
\textbf{5O-1A}  & 1.52         & .22          & 4.49         & .81         & 6.99         & 1.94        & 28.38       & 5.14       \\
\textbf{5O-2A}  & 3.50         & 1.38         & 4.63         & .62         & 8.96         & 4.69        & 20.56       & 4.60       \\
\textbf{5O-5A}  & 2.53         & .36          & 5.36         & .44         & 15.77        & 2.94        & 58.22       & 47.40      \\
\textbf{10O-5A} & 2.31         & .36          & 5.83         & 1.19        & 13.90        & 3.57        & 40.60       & 12.21     \\
\bottomrule
\end{tabular}
\end{table*}

Across all five scenarios, ChatGPT (gpt-4.1-mini) consistently exhibited the lowest latency, with mean response times ranging from .79 seconds in the simplest configuration (\textbf{1O-1A}) to around 3.50 seconds in the slightly more demanding multi-agent setup (\textbf{5O-2A}), and remaining close to 2.00-2.50 seconds in the higher-object-count scenarios (\textbf{5O-5A} and \textbf{10O-5A}). Claude (claude-sonnet-4-5) formed a second, slightly slower tier, with mean response times ranging from 3.27 seconds to 5.83 seconds across scenarios and a generally gradual increase as complexity rose.

Gemini (gemini-2.5-flash) consistently lagged behind ChatGPT and Claude, with mean response times ranging from 2.94 to 15.77 seconds, showing a clear upward trend with increasing scene complexity, particularly in multi-agent settings. Grok (grok-4-1-fast) was the slowest model across all scenarios, with average response times ranging from 4.38 seconds in the simplest case to over 58 seconds in the \textbf{5O-5A} configuration, and remaining above 20 seconds even in the intermediate-complexity conditions.

\subsection{Discussion and Limitations}
\label{sec44}
The outcomes of this work illustrate both the strengths and the boundaries of LLM-driven procedural agent-based narrative generation. One significant finding is that scene-aware prompt design, combined with explicit formatting constraints, allows even lightweight transformer models to produce highly structured outputs suitable for runtime execution. This suggests that large-scale memory, formal symbolic planners, or autonomous reasoning layers are not strictly necessary for many forms of interactive story realization when the goal is short, self-contained sequences grounded in spatial context.

Despite reliable structured output and real-time execution, the system has notable limitations in robustness, scalability, and generality. It relies on remote third-party LLM APIs, introducing network latency and dependence on external availability, pricing, and policy changes. The pipeline assumes strictly valid structured output from the LLM; occasional formatting errors can cause parsing failures, reflecting the brittleness of text-to-structure interfaces. Object conflicts may still occur in dense scenes, as conflict resolution is enforced mainly through prompt constraints rather than runtime arbitration. The virtual agents lack persistent memory, internal state, or online replanning, making the system well-suited to short, static scenarios. The action vocabulary is constrained by the available interaction types and animations, limiting narrative diversity and the ability to manipulate complex structures. Finally, system performance depends heavily on well-designed scene metadata and prompt engineering; inconsistencies can degrade both narrative coherence and execution validity.

\section{Conclusion}
\label{sec5}
This paper demonstrated that LLM-driven narrative planning can be effectively integrated with agent-based behavior orchestration in a modular 3D environment without manual scripting. By combining structured scene metadata, scene-aware prompting, and a custom \texttt{SceneDirector} parser, the system successfully translated high-level scene descriptions into executable agent behaviors in Unity. Users could author scenes through a drag-and-drop interface and automatically generate coherent procedural narratives involving navigation, animation, and object interaction. Experiments across multiple LLMs (i.e., ChatGPT, Claude, Gemini, and Grok) showed consistent production of valid, parsable action plans, with differences primarily in response time and scalability rather than correctness. Overall, the results validate the feasibility of lightweight LLM-based frameworks for rapid procedural storytelling and narrative prototyping.

Several directions can extend the framework developed in this paper. Adding agent autonomy, such as short-term memory, goal models, or reactive behaviors, could allow runtime refinement of LLM-generated plans. Incorporating dialogue, emotional modeling, and multimodal coordination would further enrich narrative expressiveness. Expanding the interaction and animation set to include physics-aware behaviors would enable more nuanced and context-sensitive actions. On the model side, integrating local or fine-tuned LLMs could reduce latency, improve determinism, and support real-time or offline use. Together, these enhancements would allow the system to handle more complex scenes and agent-based narratives, broadening its applicability across storytelling and narrative domains.

\bibliographystyle{abbrv-doi}

\bibliography{template}
\end{document}